\documentclass[a4paper,11pt]{article}

\pdfoutput=1
\usepackage{jheppub}
\usepackage[T1]{fontenc}
\usepackage{graphicx,amstext,amsmath,amssymb,epsfig,color,ulem,color}

\newcommand{\gsim}{\lower.7ex\hbox{$\;\stackrel{\textstyle>}{\sim}\;$}}
\newcommand{\lsim}{\lower.7ex\hbox{$\;\stackrel{\textstyle<}{\sim}\;$}}

\newcommand{\green}[1]{{}}

\newcommand\underrel[2]{\mathrel{\mathop{#2}\limits_{#1}}}

\newcommand{\eqn}[1]{Eq.~(\ref{#1})}
\newcommand{\fig}[1]{Fig.~(\ref{#1})}
\newcommand{\Sec}[1]{Section~\ref{#1}}
\newcommand{\Subsec}[1]{Subsection~\ref{#1}}
\newcommand{\App}[1]{Appendix~\ref{#1}}

\def\be{\begin{equation}} 
\def\ee{\end{equation}} 
\def\bea{\begin{eqnarray}} 
\def\eea{\end{eqnarray}} 
\def\del{\partial}

\def\WKBomega{\hat{\omega} \Lambda}

\def\ts{\tau_{\rm B}}
\def\harb{h}
\def\garb{g}

\def\anh{{\bar a}^{\rm NH}_0}
\def\anho{{\bar a}^{\rm NH}_1}
\def\anht{{\bar a}^{\rm NH}_2}
\def\anhth{{\bar a}^{\rm NH}_3}

\def\anb{{ a}^{\rm NB}_0}
\def\anbo{{a}^{\rm NB}_1}
\def\anbt{{ a}^{\rm NB}_2}
\def\anbth{{ a}^{\rm NB}_3}

\def\ani{{ a}^{\rm NB}_i}

\def\Danbo{\Delta{a}^{\rm NB}_1}
\def\Danbt{\Delta{ a}^{\rm NB}_2}
\def\Danbth{\Delta{ a}^{\rm NB}_3}
\def\Dani{\Delta{ a}^{\rm NB}_i}
\newcommand{\ahyd}[1]{{c}^{\rm hyd_2}_{#1}}

\def\bwkb{{b}^{\rm WKB}}
\def\awkb{{a}^{\rm WKB}}
\def\swkb{{s}^{\rm WKB}}
\def\tsb{\tau_{\rm \Omega}}
\def\bsp{{b}^{\rm SP}}
\def\asp{{a}^{\rm SP}}
\def\ssp{{s}^{\rm SP}}

\usepackage{mathtools}
\usepackage{hyperref} 
\usepackage{soul}
\hypersetup{
    unicode=true,          
    bookmarksnumbered=true,
    colorlinks=true,       
    linkcolor=red,          
    citecolor=[rgb]{0,.7,0}, 
    filecolor=magenta,
    urlcolor=cyan,
    menucolor=blue,
    baseurl=http://www.arxiv.org/abs/,
    linktocpage=true,
    hyperfootnotes=true
}

\title{\boldmath Holographic Bjorken Flow at Large-$D$}

\author{Jorge Casalderrey-Solana$^{a,b}$,}

\author{Christopher P. Herzog$^{c,d}$ and}
\author{Ben Meiring$^{b}$}
\affiliation{$^a$  Departament de F\'\i sica Qu\'antica i Astrof\'\i sica and Institut de Ci\`encies del Cosmos (ICC),\\  Universitat de Barcelona, Mart\'\i\  i Franqu\`es 1, ES-08028, Barcelona, Spain}
\affiliation{$^b$ Rudolf Peierls Centre for Theoretical Physics, University of Oxford,\\Clarendon Laboratory
Parks Road
Oxford OX1 3PU United Kingdom}
\affiliation{$^c$ Mathematics Department, King's College London, \\
The Strand, London,  WC2R 2LS, UK}
\affiliation{$^d$ C.~N.~Yang Institute for Theoretical Physics, Department of Physics and Astronomy, \\
Stony Brook University, Stony Brook, NY  11794, USA}

\emailAdd{jorge.casalderrey@ub.edu}
\emailAdd{christopher.herzog@kcl.ac.uk}
\emailAdd{ben.meiring@physics.ox.ac.uk}

\abstract{
We use gauge/gravity duality to study the dynamics of strongly coupled gauge theories undergoing boost invariant expansion in an arbitrary number of
space-time dimensions ($D$). By keeping the scale of the late-time energy density fixed, we explore the infinite-$D$ limit and study the first few corrections to this expansion. In agreement with other studies, we find that the large-$D$ dynamics are controlled by hydrodynamics and we use our computation to constrain the leading large-$D$ dependence of a certain combination of transport coefficients up to $6^{\rm th}$ order in gradients. Going beyond late time physics, we discuss how non-hydrodynamic modes appear in the large-$D$ expansion in the form of a trans-series in $D$, identical to the non-perturbative contributions to the gradient expansion. We discuss the consequence of this trans-series in the non-convergence of the large-$D$ expansion.
}

\begin{document} 
\maketitle
\flushbottom

% -----------------------------------------------------------------------------------------------------
% -----------------------------------------------------------------------------------------------------
% -----------------------------------------------------------------------------------------------------

\section{Introduction}
\label{Intro}
Gauge/gravity duality, also known as holography, is a powerful tool that allows us to explore the strongly-coupled regime of non-Abelian gauge theories by mapping their dynamics into classical gravity in anti-de Sitter (AdS) space. One of the areas where holography has had a larger impact is in the description of far-from-equilibrium dynamics of strongly coupled gauge theories. In this regime, traditional field theory methods provide limited information about the time-evolution of these types of systems.  In contrast, by now standard holographic tools are straightforward to apply, if not to QCD itself, then at least to related theories, such as maximally supersymmetric Yang-Mills.  Because of the availability of these tools, holography has had a  significant impact on physical systems in which strongly time dependent and non-equilibrium behavior are of particular relevance, such as heavy ion collisions at ultra-relativistic energies (see \cite{CasalderreySolana:2011us} for a review) or strongly correlated systems in condensed matter (see \cite{Hartnoll:2016apf} for a review).

While a gravity dual replaces a path integral description of a strongly interacting field theory with a set of nonlinear, partial differential equations, i.e.\ Einstein's equations, Einstein's equations are still difficult to solve.  
In cases of high symmetry, the equations can sometimes be solved analytically, but
the lack of symmetry in strongly time dependent and out-of-equilibrium states typically necessitates solving Einstein's equations numerically.
 (See \cite{Chesler:2013lia} for a review of numerical relativity techniques applied to holography.)  
  Indeed, solving Einstein's equations in situations with little or no symmetry is a common problem in general relativity, for example, in studying the emission of gravitational waves from black-hole mergers.
 As a way to obtain a better analytic understanding of these and other non-linear phenomena in general relativity, in recent years a new strategy based on using the number of space-time dimensions ($D$) as an expansion parameter has been developed \cite{Emparan:2013moa,Emparan:2013xia}. 
 
Since the development of this new scheme, the large-$D$ expansion has been extensively applied to different problems in holography (see e.g.\ \cite{Emparan:2013oza,Romero-Bermudez:2015bma,Emparan:2015rva,Andrade:2015hpa,
Herzog:2016hob,Rozali:2017bll,Emparan:2016sjk,Iizuka:2018zgt,Tanabe:2015hda,Bhattacharyya:2015fdk,
Tanabe:2016opw,Rozali:2016yhw,Dandekar:2016fvw,Dandekar:2016jrp,Bhattacharyya:2016nhn,
Chen:2017wpf,Chen:2017hwm,Bhattacharyya:2017hpj,Miyamoto:2017ozn,Chen:2017rxa,Herzog:2017qwp,
Dandekar:2017aiv,Emparan:2018bmi,Andrade:2018zeb,Andrade:2018nsz}). 
One common feature of most of these analyses is that they focus on dynamical situations with parametrically small (in $D$) energy fluxes in the dual field theory. This translates to parametrically small velocity fields in the hydrodynamic limit. In these situations, the large-$D$ expansion has been shown to be identical to second order hydrodynamics \cite{Emparan:2015rva, Herzog:2016hob,Rozali:2017bll,Andrade:2018zeb}, with higher viscous terms relegated to subleading $D$-corrections. In this paper we will apply this new expansion technique to a purely relativistic set-up, the dynamics of a strongly coupled field theory in $n=D-1$ space-time dimensions which experiences a boost invariant expansion, whose hydrodynamic limit is known as Bjorken flow \cite{Bjorken:1982qr}. 

Boost invariant dynamics of strongly coupled systems are of particular phenomenological importance. At least close to the central region, the matter created in the ultra-relativistic heavy ion collisions studied both at RHIC and LHC, behaves as an approximately boost invariant fluid. Most of the available hydrodynamic solvers that describe heavy ion data are based on Bjorken dynamics (or modifications of it). Beyond its phenomenological interest, Bjorken flow is also an interesting laboratory with which to explore the far-from-equilibrium dynamics of expanding matter. In the absence of transverse expansion, this is a highly symmetric configuration in which the deviation from equilibrium is controlled by a single variable, the proper time $\tau$. The holographic dual of this flow, first described in \cite{Janik:2005zt}, has allowed a detailed analysis of the 
non-convergence of the hydrodynamic expansion by analysing the behaviour of large orders in the gradient expansion \cite{Heller:2013fn}. These analyses have also led to the introduction of the concept of hydrodynamic attractors \cite{Heller:2015dha,Florkowski:2017olj}, which have been later analysed in different strongly and weakly interacting theories in the same boost-invariant setup \cite{Romatschke:2017vte,Denicol:2017lxn,Strickland:2017kux,Spalinski:2017mel,Romatschke:2017acs,Heller:2016rtz,Casalderrey-Solana:2017zyh,Behtash:2018moe,Denicol:2018pak,Heller:2018qvh,Behtash:2017wqg}. 

Building upon these advances, in this paper we will use this set-up to better understand the nature of the large-$D$ expansion.  We provide the holographic dual to Bjorken flow up to and including order $1/D^3$ terms in the perturbative expansion.  Moreover, we look at non-perturbative corrections, in the form of trans-series, at leading order in the $1/D$ expansion.  Our analysis demonstrates that the $1/D$ expansion is asymptotic and also gives insight into the early time physics, which is missing from the standard hydrodynamic approach.

The paper is organised as follows: in  \Sec{sec:HydPrem} we describe Bjorken-flow for second order hydrodynamics in an arbitrary number of dimensions, $n$,  and explore its large-$D$ limit. In \Sec{sec:LDG} we describe the holographic set-up for arbitrary space-time dimension in the gravity dual $D$. Since we are describing intrinsically relativistic dynamics, we will have to tailor the way the large-$D$ limit is taken to our problem, as we describe in \Subsec{subsec:NHR}. In \Subsec{subsec:LTE} we will compare the large-$D$ result with the late-time fluid gravity expansion of holographic Bjorken flow in arbitrary dimension. In \Subsec{subsec:BM} we describe how to relate the near-horizon fields to the dual field theory.  In \Sec{sec:NHM} we describe how non-perturbative fluctuations in the $1/D$ expansion emerge in this boost-invariant setup. The near boundary and near horizon analyses are presented in \Subsec{subsec:transNBR} and \Subsec{subsec:transNHR} respectively and the matching between the two is made in \Subsec{subsec:matching}. In \Subsec{subsec:hydrotrans} we connect these non-perturbative contributions to physics beyond hydrodynamics in the expanding plasma and non-hydrodynamical quasi-normal modes. 
Finally, in \Sec{sec:discussion} we comment on the relationship between the large-$D$ and hydrodynamic expansions in this setup as well as on the implications our findings have on the non-convergence of the large-$D$ expansion.

\section{\label{sec:HydPrem}Hydrodynamic Preliminaries}

Bjorken flow is the solution of relativistic hydrodynamics under the assumption that the fluid motion is invariant under boosts along one of the space directions of the system, $x_\parallel$. A convenient set of coordinates to express this flow is the Milne-type coordinates
\be
\label{eq:vardef}
\tau=\sqrt{t^2-x_{\parallel}^2} \,, \quad y={\rm arctanh}   \left(\frac{x_{\parallel}}{t}\right) \,,
\ee
with $\tau$ the proper time and $y$ the rapidity. As is well known, boosts along the $x_{\parallel}$-direction leave $\tau$ invariant, while $y$ transforms additively. Therefore, boost invariance translates into the independence of physical quantities on $y$. 
Assuming that the fluid has no dynamics in the hyperplane transverse to $x_\parallel$, physical properties solely depend on $\tau$, making the motion effectively one dimensional. With this assumption, independently of the number of space-time dimensions of the theory $n$, the conservation of energy and momentum equation\footnote{Note that in Milne coordinates, the Minkowski metric is non-trivial, $ds^2_{\rm Minkowski}=-d\tau^2 + \tau^2 dy^2$.} $\left . T^{\mu \nu} \right ._{; \nu}=0$  reads
\be
\label{eq:consvT}
\tau \dot \epsilon (\tau) + \epsilon(\tau) + P_L=0 \,,
\ee
where we have defined the proper energy density $\epsilon(\tau)\equiv T^{\tau \tau}$ and the longitudinal pressure $P_L\equiv\ T^y_y$.

While the conservation equation~(\ref{eq:consvT}) must be satisfied in any interacting theory (without external sources), the hydrodynamic limit provides a tremendous simplification of the dynamics, by expressing $P_L$ in a gradient expansion around the equilibrium pressure $P_{\rm eq}$.
For a conformal theory, in particular, we can the express the longitudinal pressure as 
\be
\label{eq:DeltaPLdef}
 P_L=\frac{1}{n-1} \epsilon(\tau) + \Delta P_L \, , 
\ee
where we have used the conformal equation of state $P_{\rm eq}= \epsilon_{\rm eq}/(n-1)$ and $\Delta P_L$ encodes all the deviation from the equilibrium pressure, which in the hydrodynamic limit is controlled by viscous corrections.  These corrections are determined by the viscous tensor $\Delta P_L=  \Pi^y_y$ which, 
for 1+1 dimensional expansion and to second order in gradients, is \cite{Baier:2007ix}
\be
\label{eq:Pigen}
\Pi^{\mu\nu}=-\eta \sigma^{\mu\nu} + \eta \tau_\Pi \left(D_u \sigma^{\mu \nu} + \frac{\nabla \cdot u}{n-1} \sigma^{\mu \nu} \right) + 
\lambda_1  \sigma^{\langle \mu}\,_\lambda \sigma^{\nu \rangle \lambda} \, ,
\ee
where $\eta$ is the shear viscosity, $\tau_\Pi$ and $\lambda_1$ are second order  transport coefficients, $\sigma^{\mu\nu}$ is the shear tensor, $D_u$ is the (covariant) derivative
 along the fluid velocity $u^\mu$, $D_u X \equiv u^\mu X_{;\mu}$,
 $\nabla$ the projection of the (covariant) derivative to the fluid rest frame, $\nabla^\mu X\equiv \Delta^{\mu \nu} X_{;\nu}$,
  and the brackets denote symmetrization and tracelessness in the rest fluid rest frame, $2 A^{<\mu \nu>}= \Delta^{\mu \alpha} \Delta^{\nu \beta}\left(A^{\alpha \beta} + A^{\beta \alpha} -\frac{2}{n-1}g^{\alpha\beta} A^\rho_\rho\right)$, with $\Delta^{\mu \nu}\equiv \left(g^{\mu \nu}+ u^\mu u^\nu \right)$ . 
  
Since for Bjorken flow, boost invariance imposes that the fluid velocity is fixed by symmetry $u^\tau=1$, each order in gradients entering in 
\eqn{eq:Pigen} leads to an additional inverse power of $\tau$. After evaluating the different gradient structures in \eqn{eq:Pigen}, the equation for Bjorken flow  in $n$ space-time dimensions and up to second order in gradients is \cite{Baier:2007ix}
\begin{equation}
\label{eq:Bjgen}
\tau  \frac{\dot \epsilon}{\epsilon} + \frac{n}{n-1}   - 2 \nu  \frac{1}{\tau} \frac{\eta}{\epsilon} - 2 \nu^2 \frac{\eta \tau_\pi}{\tau^2 \epsilon} 
+ 4 \nu^2  \frac{n-3}{n-2} \frac{\lambda_1}{\tau^2 \epsilon}=0\,,
\end{equation}
with $\nu=(n-2)/(n-1)$.

We would like to understand the limit of the hydrodynamic equations when the number of space-time dimensions is large. To be able to take this limit, we need additional information about how the different transport coefficients and thermodynamic properties scale with the number of space-time dimensions. This information can only be obtained from a microscopic calculation. To make connection with the holographic analysis that we will perform in the following sections, we focus on the properties of the field theory dual to a black-brane in $AdS_D$, with $D=n+1$. For this strongly coupled theory, the equilibrium energy density is given by $\epsilon_{\rm eq}=  g_*\tilde \epsilon_{\rm eq}$  with  
\be
\label{eq:etoTrel}
\tilde \epsilon_{\rm eq} =  \left( \frac{4 \pi T}{n}\right)^n\,.
\ee
and  $g_*=\frac{1}{2 \kappa^2_D} \frac{n-1}{n} $, where $\kappa_D$ is the effective $D$-dimensional Newton constant, which is related to the number of degrees of freedom of the dual gauge theory. 
Since $g_*$ is a multiplicative factor, we can re-define all thermodynamic properties analogously ($s=g_*\tilde s$, $p=g_* \tilde p$\,, ...).
Note that in the large-$D$ limit there is a parametric difference between the scale $T$ and the scale $\tilde \epsilon^{1/n}=4\pi T/n$. 
The relevant first and second order transport coefficients in this case are also known. The shear viscosity to entropy density ratio is the universal value $\eta/s=1/4\pi$ \cite{Kovtun:2004de}, while the second order coefficients are given by\footnote{Note that we have normalised the shear tensors as in  \cite{Baier:2007ix} which differs by a factor of 2 from the normalisation of \cite{Bhattacharyya:2008mz}} \cite{Bhattacharyya:2008mz}
\be
\label{eq:tcoefgen}
 \lambda_1=\frac{n}{8 \pi} \frac{\eta}{T}\, ,\quad\eta \tau_\Pi=2 \lambda_1 \left(1-\beta\right)\, ,  \quad \beta=\int_1^\infty dx \frac{x^{(n-2)}-1}{x (x^n-1)} \,.
\ee
Having specified all transport coefficients we can explicitly write the Bjorken flow equation~(\ref{eq:Bjgen}) for the strongly-coupled holographic theory. Since the hydrodynamics of a conformal theory is scale invariant, we choose to write the equation in terms of the dimensionless gradient $w_\epsilon\equiv \tau \tilde \epsilon^{1/n}$. This is analogous to the variable $w\equiv\tau T$ introduced in \cite{Heller:2011ju}; however, from \eqn{eq:etoTrel}, these two variables differ parametrically in $n$, which will be important in taking the large-$D$ limit. After some trivial thermodynamic manipulations,   \eqn{eq:Bjgen} may be written as
\begin{equation}
\label{eq:hydwegen}
\tau  \frac{\dot{ \tilde \epsilon}}{\tilde \epsilon} + \frac{n}{n-1} - \frac{2 \nu}{n-1}   \frac{1 }{w_\epsilon}  + 
 \frac{2 \nu^2}{n-1} \left( \beta - \frac{1}{n-2}
\right) \frac{1}{w_\epsilon^2} =0 \,.
\end{equation}
By comparing this expression with the conservation equation~(\ref{eq:consvT}), boost invariant flow consistent with a truncation to second order in inverse normalized gradients is given by the ratio 
\be
\label{eq:assymhyd2}
\frac{\Delta P_{\rm L}}{\epsilon} \approx \left. \frac{\Delta P_{\rm L}}{\epsilon}\right |_{\rm hyd_2} \equiv  - \frac{2 \nu}{n-1}   \frac{1 }{w_\epsilon}  + 
 \frac{2 \nu^2}{n-1} \left( \beta - \frac{1}{n-2}
\right) \frac{1}{w_\epsilon^2}  \,.
\ee
Hydrodynamics beyond second order in gradients leads to additional inverse powers of $w_\epsilon$.

At late times $w_\epsilon\rightarrow \infty$, all gradient terms in \eqn{eq:hydwegen} vanish and the dynamics are controlled by ideal Bjorken expansion
\be
\label{eq:hydpre0}
\tilde \epsilon_{\rm ideal} =   \frac{\Lambda^n}{\left(\Lambda \tau\right)^{n/(n-1)}} \,,
\ee
where we have introduced the mass-scale $\Lambda$ to take care of dimensions. At late times, all information about the initial conditions of this highly symmetric expansion is condensed in the value of this constant. When taking the large-$D$ limit, we will compare different values of the space-time dimension by fixing this late time scale, which from \eqn{eq:hydpre0} may be defined as 
\be
\label{eq:Lambdadef}
\Lambda=\lim_{\tau \rightarrow \infty}  \tau^{\frac{1}{n-2}}  \tilde  \epsilon (\tau )^{\frac{1}{n \nu}} \,,
\ee
with $\nu$ defined below \eqn{eq:Bjgen}.

Going beyond the ideal Bjorken solution, at late times a power series solution for the energy density in terms of inverse dimensionless gradient corrections can be found. Using the ideal behaviour 
\eqn{eq:hydpre0} to define the (late time) normalised gradient
\be
\label{eq:udef}
u\equiv \frac{1}{\tau \tilde \epsilon^{1/n}_{\rm ideal}}=\frac{ 1}{\left(\tau \Lambda \right)^\nu} \, ,
\ee
the energy density can be written as 
\be
\label{eq:gradexpe}
\tilde \epsilon\approx \
\tilde \epsilon_{\rm ideal}
 \left(1 -\frac{2}{n-1} u
+\frac{\nu}{n-1} \left(\frac{1}{n} +\beta\right) u^2
+\mathcal{O} \left(u^3\right) \,. 
\right)\,.
\ee
The power expansion above can, in principle, be extended to arbitrary order in $u$. However, since the hydrodynamic equation~(\ref{eq:hydwegen}) truncates the gradient 
corrections at second order, higher-powers in the $u$-expansion in the microscopic theory are sensitive to additional gradient corrections and they differ, in principle from those extracted from the power series analysis of \eqn{eq:hydwegen}.

We will now study the large-$D$ limit of the hydrodynamic equation~(\ref{eq:hydwegen}). 
As we have mentioned, we take the large-$D$ limit by keeping the late time scale $\Lambda$, \eqn{eq:Lambdadef}, fixed, which implies that 
in the large-$D$ limit $w_\epsilon \sim \mathcal{O} (1/n^0)$. 
Using the  expressions for the transport coefficients~(\ref{eq:tcoefgen}) up to order $n^{-3}$, the second order hydrodynamics equation~(\ref{eq:hydwegen}) becomes
\begin{equation}
\label{eq:lnhydro}
\tau  \frac{\dot {\tilde \epsilon}}{\tilde \epsilon} +1 +\frac{1}{n}+\frac{1}{n^2}+\frac{1}{n^3}- \frac{2}{n} \left(1-\frac{1}{n^2}\right)\frac{1 }{ w_\epsilon}  + 
\frac{1}{n} \left(1 -\frac{3}{n}- \frac{4+\frac{2 \pi
   ^2}{3}}{n^2} \right)\frac{1}{w_\epsilon^2}=0\,.
\end{equation}
Note that, at least to second order, all viscous corrections are subleading in $n$. This implies that in the infinite-$D$ limit, hydrodynamics is purely ideal and both  first and second order terms appear at next-to-leading order in the $1/n$ expansion.

We may now find a solution to the large-$D$ hydro equation~(\ref{eq:lnhydro}) in inverse powers of $1/n$ of the form 
\be
\tilde \epsilon=\tilde{\epsilon}^{\infty}_{\rm id}  \left(1+ \sum_{i=1}^{\infty} \frac{1}{n^i } \ahyd{i} \right) \,, 
\ee
where $\tilde{\epsilon}^{\infty}_{\rm id}$ is the $n\rightarrow \infty$ limit of the ideal Bjorken expansion, \eqn{eq:hydpre0},
\be
\label{eq:eidinf}
\tilde{\epsilon}^{\infty}_{\rm id}=  \Lambda^n \frac{1}{\Lambda \tau}\,.
\ee
Solving \eqn{eq:lnhydro} order by order in $1/n$ and imposing the definition~(\ref{eq:Lambdadef}) we find
\bea
\ahyd{1}&=&\frac{1}{2 \Lambda ^2 \tau
   ^2}-\frac{2}{\Lambda  \tau }-\log
   (\Lambda  \tau ) \,,
\label{eq:ahyd21}
   \\
 \ahyd{2}&=&  
   \frac{1}{8 \Lambda ^4 \tau^4}
    -\frac{1}{\Lambda ^3 \tau^3}
   +\frac{\frac{1}{2} \log (\Lambda  \tau )+1}{\Lambda ^2 \tau ^2}
   -\frac{2}{\Lambda  \tau}+\frac{1}{2} (\log (\Lambda   \tau )-2) \log (\Lambda  \tau ) \,, 
\label{eq:ahyd22}
   \\
   \ahyd{3}&=& \frac{1}{48 \Lambda ^6 \tau^6}
   -\frac{1}{4 \Lambda ^5 \tau^5}
   +\frac{\frac{3}{8} \log(\Lambda  \tau )+\frac{1}{4}}{\Lambda ^4 \tau^4}+
   \frac{\frac{4}{3}-2 \log(\Lambda  \tau )}{\Lambda ^3 \tau^3}\,,
   \label{eq:ahyd23}
   \\
   &&
   +\frac{3 \log (\Lambda  \tau )(\log (\Lambda  \tau )+6)-4 \pi^2-6}{12 \Lambda ^2 \tau^2}-\frac{2}{\Lambda  \tau}-
   \frac{1}{6} \log (\Lambda  \tau) ((\log (\Lambda  \tau )-6) \log(\Lambda  \tau )+6)\, . \nonumber
\eea
In these expressions, the inverse powers of $\Lambda \tau$ arise from the leading order expansion (in $n$) of the corresponding power of the normalised gradient $u$.
The logarithmic corrections in $\tau$ arise from the $n$-expansion of $u$. 
The next-to-leading order term, $\ahyd{1}$, agrees with the large-$n$ limit of \eqn{eq:gradexpe}, which shows that at this order, the large-$n$ and the hydro expansions coincide. 
However, subleading $n$-powers contain contributions from increasing $u$-powers. As discussed above, going beyond second order hydrodynamics, these corrections could be sensitive to higher-order gradient corrections to the stress-tensor.  In the next section, we will compute these corrections from a microscopic calculation to address how these hydrodynamic expectations compare with the full dynamics of the strongly coupled holographic theory. 

To close this section and for future notational convenience, let us briefly discuss how to go beyond second order hydrodynamics to an arbitrary order in the gradient expansion. 
 In this case, we can present the conservation equation~(\ref{eq:hydwegen}) in the form
\be
\label{thetadef}
\tau \frac{\dot {\tilde \epsilon}}{\tilde \epsilon} + \frac{n}{n-1} + \sum_{i=1}^\infty \frac{\theta^{(i)}}{w_\epsilon^j} = 0 \ ,
\ee
where the $\theta^{(i)}$ are the linear combinations, appropriate for Bjorken flow, of the $j$-th order transport coefficients.
In the large-$D$ limit, these transport coefficients can be further expanded in inverse powers of $n$:
\be
\theta^{(i)} = \sum_j \theta_j^{(i)} \frac{1}{n^j} \ .
\ee
An interesting facet of the holographic analysis to come is that, at least for $i \leq 6$, this sum over inverse powers in $n$ starts with $j = \lfloor \frac{i+1}{2} \rfloor$.

\section{\label{sec:LDG}Large-$D$ limit of Bjorken Flow} \label{sec:perturbative}

In holography, the analysis of Bjorken flow in $n$ space-time dimensions is performed  by searching for solutions of the Einstein equations in 
$D=n+1$ dimensions in an asymptotically $AdS_D$ space-time. In $D=5$ dimensions, there is an extensive literature analysing different aspects of the off-equilibrium dynamics of strongly coupled $\mathcal{N}=4$ SYM. Following many of those analyses, we will parameterize the dual space-time to Bjorken flow with the following Eddington-Finkelstein-like ansatz,
\be
ds^2 = - A(r, \,\tau) \, d\tau^2 + 2 d\tau dr + S(r, \, \tau )^2 \left( e^{-(n-2)B(r, \, \tau )} d y^2 + e^{B(r,\, \tau)} dx_{\perp}^2 \right)~,
\label{eq:metric}
\ee
where boost invariance imposes that the metric functions do not depend on the space-time rapidity $y$. As in the previous section, we will assume that there are no dynamics in the transverse plane. 
While this gauge cannot describe the dual space-time at $\tau=0$ \cite{Heller:2012je}, it is a convenient form to describe the dynamics for any $\tau>0$ since, in this form, Einstein's equations\footnote{%
In more familiar notation, the equations we solve are $R_{\mu\nu} = -(D-1) g_{\mu\nu}$ with the cosmological constant set such that an anti-de Sitter solution has radius of curvature equal to one. \green{[[ added ]]} }
 take a particularly simple sequential form \cite{Chesler:2009cy,Chesler:2013lia}, which for  $D=n+1$ dimensions are  
\bea
 S'' & =& -\frac{n-2}{4} S \left(B'\right)^2 \, , \label{eq:CY1}\\
 S \dot{S}' & =& \frac{n}{2} S^2- (n-2)\dot{S} S' \, , \label{eq:CY2} \\
 S \dot{B}' & =& -\frac{n-1}{2} \left( \dot{S} B'+\dot{B} S'\right) \, , \label{eq:CY3} \\
 A'' & = &- n (n - 3) - (n-2)(n-1)\left( \frac{1}{2}\dot{B} B' - 2 \frac{\dot{S} S'}{S^2} \right) \, , \label{eq:CY4} \\
 \label{eq:Bconstraint}
 \ddot{S} & =  &\frac{1}{2}\dot{S} A'-\frac{n-2}{4}\dot{B}^2 S  \, , \label{eq:CY5}
\eea
where $f' = \partial_r f$ and $\dot{f} = \left(\partial_{\tau} + \frac{1}{2} A(r,\tau) \partial_{r} \right) f$. 
Note that these two derivative operators do not commute, and that for any metric function $X$ we denote $\dot{X}'\equiv \left(\dot{X}\right)'$, i.e.\ the $r$-derivative here is always applied last.  
 In the characteristic formulation, the first four equations are dynamical in the bulk and can be used to solve sequentially for the different metric functions while \eqn{eq:Bconstraint} reduces to an equation for the boundary values of the functions $A$ and $B$ \cite{Chesler:2013lia}.

With the ansatz~(\ref{eq:metric}), $AdS_D$-space -- dual to the vacuum of the gauge theory -- takes a non-trivial form, which depends explicitly on proper time, $\tau$. The three metric functions in this case are given by 
\be
\label{eq:vacMF}
A_{\rm V}= r^2, \quad B_{\rm V}= \frac{2}{n-1} \log{\left(\frac{r}{1+ r \tau}\right)} , \quad S_{\rm V}=r^{\frac{n-2}{n-1}}\left(1 + r \tau \right)^{\frac{1}{n-1}}\,. 
\ee
To simplify the holographic analysis of the strongly coupled expansion, we find it convenient to factor out this non-trivial dependence of the metric functions which do not lead to dynamics in the dual field theory. Consequently, we redefine the fields as
\be
\tilde A\equiv \frac{A}{A_{\rm V}},  \quad \tilde B\equiv (n-2) \left(B - B_{\rm V}\right), \quad \tilde S\equiv \frac{S}{S_{\rm V}}. 
\ee
 This redefinition simplifies, in particular, the near boundary behaviour of the different metric functions. Exploiting the residual gauge dependence in \eqn{eq:metric}, it is always possible to choose a gauge in which close to the $AdS_D$ boundary $r\rightarrow \infty$
\bea
\label{eq:boundaryexpA}
\tilde A& =& 1 - \frac{1}{r^n} \left(\tilde \epsilon (\tau) + \mathcal{O} \left(\frac{1}{r}\right)\right) \,,
\\
\label{eq:boundaryexpB}
\tilde B& =&  \frac{1}{r^n} \left( -\frac{n}{n-1} \Delta \tilde{P}_L (\tau) + \mathcal{O} \left(\frac{1}{r}\right)\right) \, ,
\\
\label{eq:boundaryexpS}
\tilde S& =& 1+ \frac{1}{r^{n+1}} \left( s_1(\tau) + \mathcal{O}\left(\frac{1}{r}\right) \right) \ ,
\eea
where $\tilde \epsilon$ and $ \Delta \tilde{P}$ are the energy density and pressure anisotropy, \eqn{eq:DeltaPLdef}, of the dual gauge theory. 
Therefore, these two boundary values are not independent, since they are related via the conservation equation~(\ref{eq:consvT})  and \eqn{eq:DeltaPLdef}, which in holography arises from the constraints imposed by \eqn{eq:Bconstraint}. 

For large numbers of space-time dimensions, the quickly changing function $r^{-n}$ in the boundary expansion \eqn{eq:boundaryexpA} reveals two distinct regions in the gravitational analysis. Using the definition~(\ref{eq:Lambdadef}) of the scale $\Lambda$, these two regions are
\begin{itemize}
\item The Near Horizon Region:  $\log r/\Lambda \ll 1$. In this region, in the large-$n$ limit, the factor $\tilde \epsilon(\tau)/r^n$ becomes large and the gravitational field becomes strong. As expected from the presence of a non-zero density in the boundary theory, in this region the gravity dual develops a horizon at $r^n\sim 1/\tilde \epsilon$. 
\item The Near Boundary Region:  $\left(r/\Lambda\right)^n\gg 1$. In this region, the combination $\tilde \epsilon(\tau)/r^n$ is small and the effect of the energy density may be considered as a small perturbation over the $AdS_D$ space.
\end{itemize}

The treatment of these two regions requires different approximations, which can be matched at the intermediate region $1/n\ll \log r/\Lambda \ll 1$, where both the regions defined above overlap. In the following subsection we will analyse the gravitational dynamics in those two regions. 

\subsection{\label{subsec:NHR}The Near horizon region}

We start the analysis of the dual geometry for the region close to the horizon, where gravity interacts strongly and it is necessary to find a non-linear solution of Einstein's equations. This is a complicated task, which for finite $D$ in general, and for Bjorken flow, in particular, demands numerical solutions. However, as in other analyses \cite{Emparan:2013oza,Romero-Bermudez:2015bma,Emparan:2015rva,Andrade:2015hpa,
Herzog:2016hob,Rozali:2017bll,Emparan:2016sjk,Iizuka:2018zgt,Tanabe:2015hda,Bhattacharyya:2015fdk,
Tanabe:2016opw,Rozali:2016yhw,Dandekar:2016fvw,Dandekar:2016jrp,Bhattacharyya:2016nhn,
Chen:2017wpf,Chen:2017hwm,Bhattacharyya:2017hpj,Miyamoto:2017ozn,Chen:2017rxa,Herzog:2017qwp,
Dandekar:2017aiv,Emparan:2018bmi,Andrade:2018zeb,Andrade:2018nsz}, in the limit of large-$D$ Einstein's equations simplify tremendously and analytic solutions can be found.

To take the large-$D$ limit in the near horizon region, it is convenient to introduce the variable \cite{Emparan:2015rva}
\be
\label{eq:Rdef}
R=\left(\frac{r}{\Lambda}\right)^n \,,
\ee
which simplifies the analysis of the equations of motion. In terms of this variable, the intermediate region where we will perform the matching with the Near Boundary region occurs at $R\gg 1$. We may now find a solution for the different metric functions as an expansion in inverse powers of $1/n$ of the form
\bea
\tilde A(\tau,R)&=& \sum_{i=0}^{\infty} \frac{1}{n^i} A_i(\tau, R)\, , \label{eq:Ansum} \\
\tilde B(\tau,R)&=&\sum_{i=0}^{\infty} \frac{1}{n^i} B_i(\tau, R) \, , \label{eq:Bnsum} \\
\tilde S(\tau,R)&=&\sum_{i=0}^{\infty} \frac{1}{n^i} S_i(\tau, R) \, . \label{eq:Snsum} 
\eea

Before we evaluate the different functions in this expansion we would like to remark on a difference in the way the large-$D$ limit is taken in our analysis, with respect to other  computations (see e.g.\ \cite{Herzog:2016hob,Rozali:2017bll,Emparan:2016sjk,Emparan:2018bmi,Andrade:2018zeb}). Unlike those computations, in our ansatz~(\ref{eq:metric}) we do not scale the space-components of the metric differently than the time ($\tau$) or holographic component ($r$). The reason is that by doing such scaling, other analyses focus on dynamical situations in which the velocity associated with energy flow is small (of order $\mathcal{O} (1/\sqrt{n})$). This is clearly not a good approximation for Bjorken flow, since in this case the rapidity of the flow velocity can take any arbitrarily high value. However, in our scaling, which is tailored for this set-up, the effect of the pressure and viscous correction occur only at subleading order in $1/n$.

Plugging the above expansion into Einstein's equations we can determine the different metric coefficients order by order in the $n$-expansion. As already anticipated, the large-$D$ approximation simplifies the $R$-dependence of the corresponding differential equations, which can be solved analytically. 
Imposing consistency with the near boundary expansion~(\ref{eq:boundaryexpA})-(\ref{eq:boundaryexpS}),
to leading order in $1/n$ we find  
\bea
 \label{eq:largedSoln0}
A_0 & = &1 -\frac{1}{R} \anh \, , \\
B_0 & = &0 \, ,\\
S_0 & = &1 \, , 
\eea
where $\anh$ satisfies the differential equation 
\be
\label{eq:anh}
\left(1+\tau \Lambda \right) \del_{\tau } \anh - \Lambda \anh=0 \,, 
\ee
which has a simple solution as 
\be
\anh=\frac{\alpha_0}{1+\tau \Lambda}\,, 
\ee
where $\alpha_0$ is an arbitrary constant. 

The similarity between the leading order in the near boundary expansion~(\ref{eq:boundaryexpA}) and the leading order solution~(\ref{eq:largedSoln0}) suggests that the function $\anh$ may be identified, up to the scale prefactor $\Lambda^n$ which comes from the definition of $R$, with the leading order energy density, $\tilde \epsilon$. 
 However, this identification is incorrect, as may be inferred from \eqn{eq:anh}, which differs from the leading order hydrodynamic equation~(\ref{eq:lnhydro}), which to leading order in $n$ is identical to the conservation equation~(\ref{eq:consvT}).  As we will see in \Subsec{subsec:BM}, the characteristic $\left(1+\tau \Lambda \right)$ dependence of $\anh$ 
does not imply that energy density in the dual field theory at this order in $n$ possesses an infinite series in gradients $1/\tau$ at late times. This dependence is a property of the near horizon limit in the large-$D$ expansion which differs from that of the energy density. However, the late time, leading $1/\tau$ dependence agrees with the behaviour of the boundary energy density. 
Anticipating this result, and in order to lighten notation, the definition~(\ref{eq:Lambdadef}) of the scale $\Lambda$ imposes 
\be
\alpha_0=1 \, .
\ee

Following the same procedure above, the next order in the $1/n$-expansion is given by 
\bea
 \label{eq:largedSoln1}
A_1 & =& - \frac{\anho(\tau)}{R} - \frac{1}{(1+\tau \Lambda)^2} \frac{\log{R}}{R}\, , \\
B_1 & = &0 \,,  \\
S_1 & =& 0 \, ,
\eea
with 
\be
\anho=-\frac{ \log{(1+\tau \Lambda) + \alpha_1}}{(1+\tau \Lambda)}
\ee
where $\alpha_1$ is an arbitrary constant, which we fix to $\alpha_1=0$ by demanding that the overall scale of the energy density $\Lambda$ does not receive corrections in $1/n$.
 Similarly, to second order we find, 
\bea
\label{eq:largedSoln2}
A_2 & = &-\frac{\anht(\tau) }{R} + \frac{(\tau \Lambda+(1+\tau \Lambda)\log{(1+\tau \Lambda)}) }{(1+\tau \Lambda)^3}\frac{\log{R}}{R} + \frac{\tau \Lambda-1}{2 (1+ \tau \Lambda)^3} \frac{(\log{R})^2}{R}
 , \\
B_2 & = &\frac{\tau \Lambda \left(\log{\left(1+ \frac{1}{(1+\tau \Lambda) R - 1}\right)} \log{\left(\frac{1}{(1+\tau \Lambda) R}\right)} -  \text{Li}_{2}\left(\frac{1}{(1+\tau \Lambda) R}\right) \right)}{(1+\tau \Lambda)^2 \log{(1+\tau \Lambda)}} \, ,
\\
S_2 & = &0 \, ,
\eea
where $\text{Li}_{2} (z)$ is the polylogarithm  of order 2 and  
\be
\label{eqn:NHnormmodn2}
\anht(\tau) =  \frac{1}{2 (1+\tau \Lambda)^3} \left(  (1+\tau \Lambda)^2 (\log{(1+\tau \Lambda)})^2 - 2(2+4\tau \Lambda +\left(\tau \Lambda\right)^2) \log{(1+\tau \Lambda)}\right) \,,
\ee
where, as before, we have set an arbitrary integration constant by demanding that there are no $n$-corrections to $\Lambda$.
 We have also analysed the expansion to third order, which yields long expressions for the different metric functions containing polylogarithms, which,  for the reader's convenience, are tabulated in Appendix~\ref{app:longexpressions}. Here we only quote the large-$R$ behaviour of the blackening function $A$, necessary to perform the matching with the near boundary region in \Subsec{subsec:BM},
which is given by,
\bea
\label{eq:DeltaA3def}
A_3&\approx&
\frac{-\anhth(\Lambda  \tau )+\frac{4 \Lambda  \tau  (\log (\Lambda  \tau
   +1)+1)}{(\Lambda  \tau +1)^4}}{R}
    \\
   &&\hspace{-0.5cm}
   +\frac{\log (R) \left(2 \Lambda  \tau  (\Lambda  \tau +6)+\log
   (\Lambda  \tau +1) \left(10 \Lambda  \tau +(\Lambda  \tau +1)^2 (-\log (\Lambda  \tau
   +1))+4\right)\right)}{2 R (\Lambda  \tau +1)^4}
   \nonumber
      \\
   &&\hspace{-0.5cm}
   -\frac{\log ^2(R) \left(\left(\Lambda ^2 \tau^2-1\right) \log (\Lambda  \tau +1)+\Lambda  \tau  (\Lambda  \tau -5)\right)}{2 R
   (\Lambda  \tau +1)^4}
   -\frac{(\Lambda  \tau  (\Lambda  \tau -4)+1) \log ^3(R)}{6 R
   (\Lambda  \tau +1)^4} \, ,
\nonumber  
\eea
with,
\bea
\label{eq:a3nh}
\anhth(\tau)&=&
 -\frac{18 \Lambda  \tau +\pi ^2 (3 \Lambda  \tau -1)+6}{9 (\Lambda  \tau +1)^4}
 -\frac{\log ^3(\Lambda 
   \tau +1)}{6 \Lambda  \tau +6}
   +\frac{(\Lambda  \tau +2) \log ^2(\Lambda  \tau +1)}{(\Lambda  \tau +1)^2} \, \nonumber \\
  & & - \frac{(\Lambda 
   \tau +2) (\Lambda  \tau  (\Lambda  \tau +5)+2) \log (\Lambda  \tau +1)}{(\Lambda  \tau +1)^4} \, .
\eea
The function $\tilde S -1$ remains zero to next-to-next-to-next-to-leading order;  we have checked that it is not zero at all orders, with the first non-trivial contribution appearing for $S_4$. Since this fourth-order expression is needed to determine $A_3$ and $B_3$, its expression can also be found in Appendix~\ref{app:longexpressions}.

As we have already mentioned, to extract the dual field theory stress tensor from these expansions, we need to perform a matching calculation of these results with the near boundary region, which we describe in \Subsec{subsec:BM}. Prior to this, in the next section we will compare these
results with the results from a gradient-expansion of the gravity dual for Bjorken flow. 

\subsection{\label{subsec:LTE}Comparison with late time solutions at large-$D$}
Since as we have seen in \Sec{sec:HydPrem} gradients in Bjorken flow decrease at late times, for large $\tau$ we can do an alternative expansion for Einstein's equations at fixed number of space-time dimensions $D=n+1$, similar to the fluid-gravity correspondence \cite{Bhattacharyya:2008jc}.
For the particular case of boost invariant dynamics in $D=5$ this analysis was first performed in \cite{Kinoshita:2008dq} and employed in \cite{Heller:2013fn} to study large orders of the gradients expansion. In this section we will generalise the analysis of \cite{Kinoshita:2008dq} to arbitary $D$ and compare the large-$D$ limit of those late time solutions to the near horizon solutions found in \Subsec{subsec:NHR}.

To make contact with the notation in \cite{Kinoshita:2008dq} and to facilitate the late-time analysis, we introduce the following field redefinition 
\bea
\label{eq:gexpgenA}
A & = & r^2 \bar{A}(\tau,r), \\
\label{eq:gexpgenB}
B & = &\frac{1}{n-1} \left( \log{\left(\frac{r^2}{(1+ r \tau)^2}\right)} +   \frac{n-1}{n-2} \left(2 \, \bar{d}(\tau,r) -\bar{B}(\tau,r) \right) \right), \\
\label{eq:gexpgenS}
S & = &r^{\frac{n-2}{n-1}} (1+ r \tau)^{\frac{1}{n-1}} e^{\bar{d}(\tau,r)}.
\eea
At late times, when hydrodynamics becomes a valid approximation, the dual gravity space-time should be very close to that of a boosted black-brane with a time-dependent energy density, predicted by  ideal hydrodynamics, $\tilde \epsilon_{\rm ideal}$ in \eqn{eq:hydpre0}. The corrections to this approximate solution are controlled by a dimensionless gradient, \eqn{eq:udef}. Generalising \cite{Kinoshita:2008dq}, it is convenient to parametrise the holographic direction in terms of a new variable that tracks the position of the boosted black-brane horizon,
\be
 s=\frac{\Lambda}{r} \frac{1}{\left(\tau \Lambda \right)^{-1/(n-1)}} \, ,
\ee
such that $s=1$ corresponds to the position of the horizon to leading order in gradients. The gradient 
expansion of the different metric functions is 
\bea
\bar{A}(r,\tau) & =& \sum_{i=0}^\infty u^i \bar{A}_i(s)~,
\label{eq:Aexp} \\
\bar{B}(r,\tau) & = &\sum_{i=0}^\infty u^i \bar{b}_i(s)~,
\label{eq:Bexp} \\
\bar{d}(r,\tau) & = &\sum_{i=0}^\infty u^i \bar{d}_i(s)~.
\label{eq:dexp}
\eea
Introducing these expansions into Einstein's equations and expanding in the dimensionless gradient, $u$, the leading order (in gradients) solution is simply the boosted black brane
\bea \label{zeroHydro}
\bar{A}_{0}(s) & =& 1 - s^{n} \, , \\
\bar{b}_{0}(s) & =& 0 \, , \\
\bar{d}_{0}(s) & =& 0 .
\eea
Going to next-to-leading order in gradients, the metric functions include the effect of the shear viscosity and they read 
\bea
\bar{A}_{1}(s) & =& \frac{(2+(n-2)s)s^n}{n-1} \, , \\
\bar{b}_{1}(s) & =& \frac{2(n-2)}{n(n-1)}\left( \beta(s^n \, ; 1+\frac{1}{n},0) + \log{(1-s^{n})}\right) \, ,\\
\bar{d}_{1}(s) & =& 0 \, ,
\eea
where $\beta(z;a,b)$ is the incomplete beta function. Note that we have chosen a gauge consistent with the near-boundary behaviours~(\ref{eq:boundaryexpA})-(\ref{eq:boundaryexpS}). We have not been able to find closed form expressions for these metric functions beyond this order. 

Inserting this gradient expansion into Eqs.\ (\ref{eq:gexpgenA})-(\ref{eq:gexpgenS}) and expressing them in $\left(\tau, \,R\right)$,
we can compare the fluid-gravity expectation for the different metric functions with the near horizon result derived in the previous section.  After taking the large-$D$ limit and expanding to $\mathcal{O}\left(n^{-3}\right)$ 

\bea
\label{eq:Athyd}
\tilde{A}_{\text{hydro}}(r,\tau) & = & \, \left(1-\left(\frac{1}{\tau \Lambda} -\frac{1}{{(\tau \Lambda)}^2}\right)\frac{1}{R}\right) 
+ \frac{1}{n} \left(\frac{1+(-1 + \tau\Lambda)\log{(\tau)}}{{(\tau \Lambda)}^2 R} - \frac{\log{R}}{(\tau\Lambda)^2 R} \right) + \\
&&  + \frac{1}{n^2} \left(
\frac{2+ 2\tau \Lambda \log{\tau \Lambda} - (-1+\tau \Lambda) \log{(\tau \Lambda})^{2}}{2 (\tau \Lambda)^2 R} 
 + \frac{(1+\log{\tau \Lambda})\log{R}}{(\tau \Lambda)^2 R} +\frac{(\log{R})^{2}}{2 (\tau \Lambda)^2 R} \right)  \, , \nonumber \\
 \label{eq:Bthyd}
  \tilde{B}_{\text{hydro}}(r,\tau) &= & \frac{1}{n^2} \left( \frac{2 \left(\text{Li}_2\left(\frac{1}{R \Lambda  \tau }\right)+\log \left(1-\frac{1}{\Lambda
    R \tau }\right) \log \left(\frac{1}{\Lambda  R \tau }\right)\right)}{\Lambda  \tau } \right) \, ,
  \\
  \label{eq:Sthyd}
   \tilde{S}_{\text{hydro}}(r,\tau) &= & \, 1 \, .
\eea
Since these solutions come from a gradient analysis, they are only valid up to $\mathcal{O}(\tau^{-2})$. These expressions coincide with the late-time expansion of  Eqs.~(\ref{eq:largedSoln0})-(\ref{eqn:NHnormmodn2}) up to order $\mathcal{O}(\tau^{-2})$, indicating, as expected from the hydrodynamic analysis of \Sec{sec:HydPrem}, that the large-$D$ expansion up to this order is controlled by hydrodynamics. Note, that in  
\eqn{eq:Athyd} and \eqn{eq:Bthyd}
 the viscous contribution, proportional to $u$,  affects both the $1/n$ and $1/n^2$ contribution, while the $1/n^0$ is completely determined by the boosted black brane contribution after taking the large-$D$ limit. The matching of these approaches may be considered as an example of the compatibility of the gradient and the large-$D$ expansions \cite{Bhattacharyya:2018iwt}.

\subsection{\label{subsec:BM}Matching with the near boundary region}

We now turn to the near boundary region, $\log r/\Lambda \gg 1$. As already mentioned, while in the near-horizon region gravity interacts strongly, in this region we can view the effect of the horizon as a small perturbation on $AdS_D$. To make the approximation apparent, we will redefine the fields in this region as 
\bea 
 A(\tau,r) &=& r^2 \left(1 - \frac{a(r,\tau)}{r^{n}} \right) \ , \label{eq:linearA} \\
B(\tau,r) &=& \frac{2}{n-1} \log{\left(\frac{r}{1+ r \tau}\right)} + \frac{1}{n-2} \frac{b(r,\tau)}{r^{n}} \ , \label{eq:linearB} \\
S(\tau,r) &=& r \left( \frac{1+r \tau}{r} \right)^{\frac{1}{n-1}} \left( 1 + \frac{s(r,\tau )}{r^{n+1}} \right) \ . \label{eq:linearS}
\eea
Since gravity is weak in this limit, we  linearize the equations dropping all terms that are quadratic or higher in $a$, $b$, and $s$.  Taking appropriate linear combinations of the equations of motion, we find the following fourth order equation for $a(r,\tau)$,
\bea
\label{eq:master}
0&=& 
-\frac{n (n+1) ((n-1) r \tau+n+2) a^{(0,1)}(r,\tau)}{r^5 (r \tau+1)}+\frac{2 n \left(2
   n+\frac{3}{r \tau+1}-2\right) a^{(1,1)}(r,\tau)}{r^4} \nonumber \\
   &&
   -\frac{(n-3) (n-1) n
   a^{(1,0)}(r,\tau)}{r^3}+\frac{\left(-5 n-\frac{3}{r \tau+1}+5\right)
   a^{(2,1)}(r,\tau)}{r^3}\nonumber \\
   &&
   +\frac{(n-1) (3 n-8) a^{(2,0)}(r,\tau)}{r^2}+\frac{(7-3 n)
   a^{(3,0)}(r,\tau)}{r}+\frac{2 a^{(3,1)}(r,\tau)}{r^2}+a^{(4,0)}(r,\tau) \, . \nonumber 
   \\
\eea
Assuming a series expansion of $a(\tau,r)$ in powers of $1/n$, we solve this equation order by order in $1/n$. From the analysis of the leading order equation, we find that it is convenient to introduce a new time variable $ \ts= \tau+ r^{-1}$. Note that this variable agrees with the gauge theory proper-time at the boundary. 
The solution up to order $1/n^3$ is
\bea
\label{eq:anps}
a(\tau,r) &=& \anb(\ts) +\frac{1}{n}\left( \anbo{}(\ts) + \Danbo{}(\ts,r) \right) 
 \\&&
+\frac{1}{n^2}\left( \anbt{}(\ts) + \Danbt{}(\ts,r) \right)
+\frac{1}{n^3}\left( \anbth{}(\ts) + \Danbth{}(\ts,r) \right)\,, \nonumber
\eea
where the functions $\ani (\ts)$ are only functions of the boundary proper time and can only be determined after matching with the near horizon behaviour, while the functions $\Dani(\ts,r)$
are fixed by the  $\ani (\ts)$ as 
\bea
\Danbo{}(\ts,r)&=& \frac{(3 - 4 r \ts) \anb{}'(\ts) + \ts \anb{}'' (\ts)}{2 r^2 \ts} \,, \\
\Danbt{}(\ts,r)&=&  \frac{-3 + 24 r \ts - 48 r^2 \ts^2 + 32 r^3 \ts^3) \anb{}'(\ts)}{8 r (r \ts-1)^3} + \frac{(3 - 4 r \ts) \anbo{}'(\ts)}{2 r^2 \ts} 
 \\
&&
\hspace{-2cm}
+ \frac{(3-24 r \ts+8 r^2 \ts^2) \anb{}''(\ts)}{8 r^4 \ts^2} + \frac{\anbo{}''(\ts)}{2 r^2} + \frac{(3 - 4 r \ts) \anb{}^{(3)}(\ts)}{4 r^4 \ts} + \frac{\anb{}^{(4)}(\ts)}{8 r^4} \,,
\nonumber
\\
\Danbth{}(\ts,r)&=&
\frac{\anb{}^{(3)}\left(\ts\right) \left(3 r \ts-1\right)}{4 r^6 \ts^3}+
\frac{\anb{}^{(5)}\left(\ts\right) \left(3-4 r \ts\right)}{16 r^6 \ts}
+\frac{\anb{}^{(6)}\left(\ts\right)}{48 r^6}
\\
&&
+\frac{\anbo{}^{(3)}\left(\ts\right) \left(3-4 r \ts\right)}{4 r^4 \ts}+
\frac{\anbo{}^{(4)}\left(\ts\right)}{8 r^4}+\frac{\left(3-4 r \ts\right) \anbt{}'\left(\ts\right)}{2 r^2
   \ts}+\frac{\anbt{}''\left(\ts\right)}{2 r^2}
 \nonumber   \\
   &&
   +
   \frac{3 \anb{}^{(4)}\left(\ts\right) \left(4 r^2 \ts^2-8 r \ts+1\right)}{16 r^6 \ts^2}
   -\frac{3
   \left(8 r^2 \ts^2-8 r \ts+1\right) \anbo{}'\left(\ts\right)}{8 r^4 \ts^3}
 \nonumber      \\
   &&
   +\frac{\left(-32 r^4 \ts^4+192 r^3 \ts^3-180 r^2 \ts^2+36 r \ts+3\right) \anb{}''\left(\ts\right)}{16 r^6 \ts^4}
 \nonumber   \\
   &&
\hspace{-2.5cm}   +\frac{\left(16 r^2 \ts^2-24 r \ts+3\right) \anbo{}''\left(\ts\right)}{8
   r^4 \ts^2}+\frac{3 \left(16 r^4 \ts^4-64 r^3 \ts^3+60 r^2 \ts^2-12 r
   \ts-1\right) \anb{}'\left(\ts\right)}{16 r^6 \ts^5} \,.\nonumber
\eea
Since the functions $\Dani(\ts, r)$ vanish at the boundary, the functions $\ani(\ts)$ correspond to the $1/n$ expansion of the coefficient $a_0(\ts)$ of the normalisable mode of the metric, \eqn{eq:boundaryexpA},
which in turn determines the $n$-expansion of the energy density $\tilde \epsilon$
as  
\be
\tilde \epsilon(\tau)=\anb(\tau) +\frac{1}{n} \anbo(\tau) + \frac{1}{n^2} \anbt(\tau) + \frac{1}{n^3} \anbth(\tau)\, .
\ee
 The solution~(\ref{eq:anps}) can be used extract the energy density in the dual theory from  the  near horizon analysis performed in \Subsec{subsec:NHR}. After replacing $r=\Lambda R^{1/n}$ in that expression and taking the large-$n$ limit, we can match the large $R$ behaviour 
of \eqn{eq:anps} with Eqs.~(\ref{eq:largedSoln0})-(\ref{eqn:NHnormmodn2}) by identifying 
\bea
\label{eq:ahol0}
\anb(\tau) &=&\tilde{\epsilon}^{\infty}_{\rm id} (\tau) \ , \\
\label{eq:ahol1}
\anbo(\tau) &=& \tilde{\epsilon}^{\infty}_{\rm id} (\tau)  \, \ahyd{1}(\tau) \,, 
\\
\label{eq:ahol2}
\anbt(\tau) &=&\tilde{\epsilon}^{\infty}_{\rm id} (\tau)  \left(
\frac{\theta _2^{\text{(3)}}}{3
   \Lambda ^3 \tau ^3}+\frac{\theta
   _2^{\text{(4)}}}{4 \Lambda ^4
   \tau ^4}
   +\ahyd{2}(\tau)
\right) \,,
\\
\label{eq:ahol3}
\anbth(\tau) &=& \tilde{\epsilon}^{\infty}_{\rm id} (\tau)
\left(
\frac{\frac{1}{6} \left(\theta _2^{\text{(3)}}-3 \theta_2^{\text{(4)}}\right)
  +\frac{\theta_3^{\text{(5)}}}{5}}{\Lambda ^5 \tau ^5}+
  \frac{-8 \theta _2^{\text{(3)}}+9 \theta _2^{\text{(4)}} \log(\Lambda  \tau )+3 \left(\theta _2^{\text{(4)}}
  +\theta_3^{\text{(4)}}\right)}{12 \Lambda ^4 \tau ^4}+ \right .
 \nonumber \\
  &&
\left. + \frac{2
   \theta _2^{\text{(3)}} \log (\Lambda  \tau )+\theta
   _2^{\text{(3)}}+ \theta_3^{(3)}}{3 \Lambda ^3 \tau
   ^3}+\frac{\frac{\theta
   _2^{\text{(4)}}}{8}+\frac{\theta
   _3^{\text{(6)}}}{6}}{\Lambda ^6 \tau ^6} + \ahyd{3}(\tau) \right) \,,
 \label{eq:athirdordern}
\eea
where $\tilde{\epsilon}^{\infty}_{\rm id} (\tau)$ is the large-$D$ limit (\ref{eq:eidinf}) of ideal Bjorken flow,
the functions $ \ahyd{1}(\tau)$, $\ahyd{2}(\tau)$ and $\ahyd{3}(\tau)$ are obtained from the second order hydrodynamics
Eqs.\ (\ref{eq:ahyd21})-(\ref{eq:ahyd23}),
and the coefficients $\theta^{(i)}_2$ and $\theta^{(i)}_3$ are transport coefficients defined in Eq.\ (\ref{thetadef}):
\bea
\theta^{(3)}_2&= 3\,, \quad &\theta^{(3)}_3=  \, \frac{-9 + 4 \pi^2}{3} \, ,
\\
\theta^{(4)}_2&=-2 \,, \quad &\theta^{(4)}_3=  \, 6 \, ,
\\
& &\theta^{(5)}_3=  \, -\frac{25}{4} \, , \\
& &\theta^{(6)}_3=  \, 7 \, .
\eea

As is apparent from Eqs.\ (\ref{eq:ahol0}) and (\ref{eq:ahol1}), our large-$D$ computation coincides exactly with the prediction of second order hydrodynamics. 
This observation implies that at this order in the $n$ expansion, all non-hydrodynamic modes are decoupled from the time evolution of the system. It also implies that 
the relevant higher-gradient transport coefficients (normalised by the energy density) grow slowly with the number of dimensions, such that at large $n$ their contribution is subleading. This result is in fact expected from the observation of other large-$D$ analyses that have observed that for small velocity set-ups, second order hydrodynamics becomes an exact description of the system dynamics \cite{Emparan:2015rva, Herzog:2016hob,Rozali:2017bll,Andrade:2018zeb}. Going beyond $1/n$-order, the system evolution is no longer described by second order hydrodynamics. Nevertheless, the difference between the microscopic calculation and the truncation of the stress tensor shown in \eqn{eq:ahol2} and \eqn{eq:ahol3} can be expressed in terms of inverse powers of the normalised gradient $w_\epsilon$. In fact, Eqs.\ (\ref{eq:ahol0})-(\ref{eq:ahol3}) may be viewed as the prediction of 6-th order hydrodynamics, for which the deviation of the longitudinal pressure with respect to its equilibrium value obtains contributions up to $w_\epsilon^{-6}$
\be
\label{eq:DeltaPextended}
\frac{\Delta P_{\rm L}}{\epsilon} \approx \left. \frac{\Delta P_{\rm L}}{\epsilon}\right |_{\rm hyd_2} +
 \sum_{i=3}^{6} \frac{\theta^{(i)}}{w^i_\epsilon} \,,
\ee
where we have used \eqn{eq:assymhyd2}.
Since no terms with inverse powers of $w_\epsilon$ greater than 6 appear in our holographic result,  \eqn{eq:ahol2} and \eqn{eq:ahol3} specify the result of this expansion up to $\mathcal{O} (n^{-4})$ for all higher order transport coefficients.

\section{Non-Perturbative Modes and Large-$D$ Trans-series }
\label{sec:NHM}

When finding the large $D$ expansion in the preceding section it was clear that the resulting perturbative solution was unique. However for finite $D$ we should expect that the metric could undergo different paths of evolution, corresponding to different choices of initial data. One could ask whether there are additional solutions to Einstein's equations which are non-perturbative in $D = n + 1$.

In the previous case we performed an expansion around an asymptotically AdS geometry in powers of inverse $D$. In this section we will study perturbations that can exist on top of this background which evolve on short timescales of order $1/n\,\Lambda$. 
As we will see, this is the typical scale of variation of the characteristic relaxation modes of black branes, known as quasi-normal modes (QNMs), which were first analysed at large-$D$ in \cite{Emparan:2014cia,Emparan:2014aba}. To describe the dynamics of those fast fluctuations we will employ WKB-like techniques, which can also be used to describe the spectrum of QNMs, as demonstrated in \App{app:QNM}.

\subsection{Non-perturbative contributions in $1/D$ and the near boundary region}
\label{subsec:transNBR} 
\green{[[ Tried to address incompatible definitions of $\hat \omega$ by introducing $\Lambda$ below via the new definition WKBomega. ]]}

 Using the linearized Einstein's equations (\ref{eq:linearA}) to (\ref{eq:linearS}), we consider the Ansatz,
\begin{align}
b(r,\tau) & =  \left( \bwkb_0(r,\tau)  + ... \right) \Omega(r,\tau) \, , \label{eq:blinear} \\
a(r,\tau) & = \left( \frac{1}{n} \awkb_1(r,\tau)  + ... \right) \Omega(r,\tau) \, , \label{eq:alinear} \\
s(r,\tau) & = \left( \frac{1}{n^2} \swkb_2(r,\tau)  + ... \right) \Omega(r,\tau) \, . \label{eq:slinear}
\end{align}
where we choose,
\begin{equation} \label{OmegaAnsatz} 
\Omega(r,\tau) = \left( \frac{r}{\Lambda} \right)^{n/2} e^{- i n \WKBomega \left(\tau +\frac{1}{r}\right) } \,e^{n \sigma(r) } \, .
\end{equation}
This form of $\Omega(r,\tau)$ 
is constrained in the large $n$ limit by time translation invariance, as we argue in Appendix~\ref{generalOmegaAnsatz}. To leading order in $n$ one finds that $\sigma'(r) = \pm\frac{\sqrt{r^2-4 (\WKBomega)^2}}{2 r^2}$.
Choosing the normalizable (``$-$'') branch of the square root we find the solutions,
\begin{align}
\sigma(r) & = \frac{1}{2} \left(\frac{\sqrt{r^2-4 (\WKBomega)^2}}{r}-\log \left( \sqrt{\frac{r^2}{\Lambda^2}-4 \hat \omega^2}+\frac{r}{\Lambda}\right)\right) \, , \\
\awkb_1(r,\tau) & = \frac{i r \left(\sqrt{r^2-4 (\WKBomega)^2}+r\right)}{2 \WKBomega (1 + r \tau) \left(\sqrt{r^2-4 (\WKBomega)^2}-2 i \WKBomega \right)} \bwkb_0(r,\tau) \, , \label{eq:Anb1} \\
\swkb_2(r,\tau)& = \frac{2 r^2}{(1 + r \tau) \left(\sqrt{r^2-4 (\WKBomega)^2}+r-2 i \WKBomega \right)} \bwkb_0(r,\tau) \, .
\end{align}
At the next order in inverse $n$ we find that we can further constrain $\bwkb_0(r,\tau)$ by,
\begin{equation}
\bwkb_0(r,\tau) = \harb_{0}(\Lambda \tsb) \frac{\Lambda^n r}{\sqrt{1+ r \tau} \sqrt[4]{\frac{r^2}{\Lambda^2}-4 \hat \omega^2}} \, ,
\end{equation}
with $\tsb = \tau+\frac{1+\frac{i \sqrt{r^2-4 (\WKBomega)^2}}{2 \WKBomega }}{r}$ and $\harb_{0}(\tsb)$ an undetermined function of $\tsb$. 

Expanding (\ref{eq:Anb1}) close to the boundary and using the relation between the near boundary metric and stress tensor \eqn{eq:boundaryexpA}, the near boundary energy density of the fluctuation is given by, 
\be
\label{eq:deltaepsilonh}
\delta \tilde \epsilon(\tau) = \frac{ i \Lambda^n }{n \hat \omega}
\left( \frac{e}{2} \right)^{\frac{n}{2}}
\frac{\harb_0\left(\tau \Lambda+\frac{i}{2 \hat \omega }\right)}{\left(\Lambda \tau\right)^{3/2}} e^{- i n \hat \omega \Lambda \tau} + \mathcal{O}(n^{-2}) \, .
\ee

The expansion of Eq.s (\ref{eq:blinear}) to (\ref{eq:slinear})  above can be viewed in analogy with Quantum Mechanics as a WKB expansion in $n$, and will not be valid for values of $r$ which correspond to 
stationary points of $\sigma(r)$, given by the condition\footnote{%
The precise condition for the validity of the WKB approximation is more complicated than this because it involves $\tau$ and $r$ derivatives of $\bwkb_0$. However, close to the stationary point, it reduces to the stated condition.} 
 $n \, \sigma'(r)^2 = \sigma''(r)$, which in this case is given by $r = 2 \WKBomega$.
It is standard in such situations to ``zoom in'' on this region by linearizing the equations of motion about the stationary point.  Near this point, the ``phase of the wave-function'' has the expansion
\be
\sigma(r) = - \frac{1}{2} \log(2\WKBomega) - \frac{(r- 2 \WKBomega)^{3/2}}{6 \WKBomega^{3/2}} + O(r- 2 \WKBomega)^{5/2} \ ,
\ee
which motivates the definition of a new coordinate,
\be
 x \equiv \left(\frac{r}{2 \hat \omega \Lambda} -1 \right)n^{2/3} \, .
\ee
In the region $\left | x \right| \ll 1$, the conditions for the validity of WKB are not met and a solution in the vicinity of $x = 0$ must be found.

Linearizing Eq.'s (\ref{eq:linearA}) to (\ref{eq:linearS}) about $x = 0$ with an Ansatz of the form,
\begin{align}
b(r,\tau) & = \left(  \bsp_0(x,\tau) + n^{-1/3} \bsp_{1}(x,\tau) + ... \right) r^{n/2} e^{- i n \WKBomega\left(\tau +\frac{1}{r}\right) } \, , \label{eq:blinearIR} \\
a(r,\tau) & = \frac{1}{n} \left(  \asp_{1}(x,\tau) + n^{-1/3} \asp_{2}(x,\tau) + ... \right) r^{n/2} e^{- i n \WKBomega\left(\tau +\frac{1}{r}\right) } \, , \label{eq:alinearIR} \\
s(r,\tau) & = \frac{1}{n^2}\left(  \ssp_{2}(x,\tau) + n^{-1/3} \ssp_{3}(x,\tau) + ... \right) r^{n/2} e^{- i n \WKBomega\left(\tau +\frac{1}{r}\right) } \, , \label{eq:slinearIR}
\end{align}
we find an Airy function equation for $\bsp_0$, and simple algebraic relations for $\asp_1$ and $\ssp_2$ that relate them to $\bsp_0$.
\begin{align}
\bsp_0(x,\tau) & = \garb_{0}(\tau) \, \text{Ai}\left(\frac{x}{\sqrt[3]{2} }\right) \, , \\
\asp_1(x,\tau) & = - \frac{\garb_{0}(\tau)}{1+2 \WKBomega \tau} \, \text{Ai}\left(\frac{x}{ \sqrt[3]{2} }\right)  \, , \\
\ssp_2(x,\tau) & = \frac{(2+2 i) \WKBomega  \garb_{0}(\tau) }{1 + 2 \WKBomega \tau} \, \text{Ai}\left(\frac{x}{ \sqrt[3]{2}  }\right) \, ,
\end{align}
where $\garb_{0}(\tau)$ is an arbitrary function of $\tau$. At subleading order we find,
\begin{equation}
\bsp_1(x,\tau) = \garb_{1}(\tau) \text{Ai}\left(\frac{x}{\sqrt[3]{2} }\right)-\frac{i \left((1+ 2 \WKBomega \tau) \garb_{0}'(\tau)+\WKBomega  \garb_{0}(\tau)\right)}{\sqrt[3]{2} \WKBomega  (1+ 2 \WKBomega \tau )} \text{Ai}'\left(\frac{x}{ \sqrt[3]{2} 
   }\right) \, .
\end{equation}
Similar algebraic equations relate $a_2^{\rm SP}(x,\tau)$ and $s_3^{\rm SP}(x,\tau)$ to $b_1^{\rm SP}(x,\tau)$, but neither will be relevant for the following discussion.  

Matching of the results of Eq.'s (\ref{eq:blinear}) and (\ref{eq:blinearIR}) in the region where $x\gg1$ at leading order in $n$ gives, 
\be
\label{eq:mnbairy}
\harb_{0}(\Lambda \tau) = \frac{
    (2\hat{\omega})^{n/2}
   }{\sqrt[6]{16 n}} \sqrt{\frac{\Lambda \tau}{\pi}} \garb_{0}\left(\Lambda \tau-\frac{1}{2
   \hat{\omega}}\right) \,.
\ee

\subsection{Non-perturbative contributions in $1/D$ and the near horizon region}
\label{subsec:transNHR} 
Informed by the structure of Eq. (\ref{OmegaAnsatz}), we make make an Ansatz in the near horizon region of the form,
\bea
\tilde A(\tau,R)&=& \sum_{i=0}^{\infty} \frac{1}{n^i} A_i(\tau, R) + \sum_j \Omega^{\rm NH}_j (\tau) \sum_{i=0}^\infty \frac{1}{n^{i}} A_{i j}(\tau, R) + ... \, , \label{eq:Anflux} \\
\tilde B(\tau,R)&=&\sum_{i=0}^{\infty} \frac{1}{n^i} B_i(\tau, R)  + \sum_j \Omega^{\rm NH}_j (\tau) \sum_{i=0}^\infty \frac{1}{n^{i}} B_{i j}(\tau, R)  + ... \, . \label{eq:Bnflux} \\
\tilde S(\tau,R)&=&\sum_{i=0}^{\infty} \frac{1}{n^i} S_i(\tau, R)  + \sum_j \Omega^{\rm NH}_j (\tau) \sum_{i=0}^\infty \frac{1}{n^{i}} S_{i j}(\tau, R)  + ... \, , \label{eq:Snflux} 
\eea
where,
\be \label{eq:OmegaNH}
\Omega^{\rm NH}_j (\tau)= e^{- i  n \hat{\omega}_{j} \left(1+\tau \Lambda\right)} \, .
\ee
Here the subscript $j$ anticipates a family of solutions parameterized by $\hat{\omega}_{j}$. The ellipses denote higher order terms in the trans-series, which arise as powers of $\Omega_{j}^{\rm NH}(\tau)$.
This Ansatz supplements the expansion in powers of $1/n$ from Eqs.\ (\ref{eq:Ansum})-(\ref{eq:Snsum}) with a non-perturbative contribution in $n$. From the linear independence of $\Omega^{\rm NH}_j (\tau)$, the functions  $A_i(\tau, R)$, $B_i(\tau, R)$ and $S_i(\tau, R)$ can be solved identically to the perturbative series 
analysed in \Subsec{subsec:NHR}. The $\mathcal{O}(\Omega^{\rm NH}_j (\tau))$ contributions given by $A_{i j}(\tau, R)$, $B_{i j}(\tau, R)$ and $S_{i j}(\tau, R)$ describe linearised fluctuations on top of the $1/D$ expanded non-linear background.
Inserting the ansatz (\ref{eq:Anflux})-(\ref{eq:Snflux}), 
we can find the first few terms of the trans-series in the near-horizon region. Many of the lowest order coefficients vanish,
\begin{align}
\label{eq:ANHnp}
  A_{0 j} & = 0  \, , \\
  S_{0 j} & = S_{1 j} = 0\,,
\end{align}
while the metric function $B_{0 j}(R,\tau) = Z(\xi,\tau)$ with $\xi = (1+\Lambda \tau) R$ satisfies,
\bea \label{eq:phase}
Z'' + \frac{(1 + 2 \xi (-1 + i \hat \omega))}{ \xi (1- \xi)} Z' + \frac{i \hat \omega}{ \xi (1- \xi)} Z&=& 0 \, ,
\eea
where $'$ denotes derivatives with respect to $\xi$. This equation coincides with the large-$D$  limit of the QNM equation of a static black-brane, which gives the near-horizon behaviour  of the spin-2, zero momentum QNM analysed in Appendix~\ref{app:QNM}.
 Imposing that $Z(\xi,\tau)$ is regular at $\xi = 1$, one can find a solution to Eq. (\ref{eq:phase}) given by,
\begin{equation} \label{eq:B0NRsoln}
	B_{0 j}(R,\tau) = f(\tau) \, \phantom{}_{2} F_{1}(q_{+},q_{-},q_{+} + q_{-}, 1- (1+ \Lambda \tau) R ) \, ,
\end{equation}
with ${}_2F_1$ the ordinary hypergeometric function,  $q_{\pm} = \frac{1}{2} \left(1-2 i \hat{\omega} \pm \sqrt{1-4 \hat{\omega}^2}\right)$ and $f(\tau)$ an arbitrary function of $\tau$.
Beyond this leading order, 
the next non-zero corrections in the large-$D$ expansion are $A_{1 j}$, $B_{1 j}$ and $S_{2 j}$, which we will not compute in this work. 

\subsection{Matching in the overlap region}
\label{subsec:matching}
Similar to the perturbative contributions discussed in \Sec{sec:LDG}, the near horizon analysis of the previous subsection must be matched to the near boundary region in order to extract the energy density associated with the corresponding  gauge theory. As in the analysis of QNMs in Appendix~\ref{app:QNM}, this matching procedure imposes that not all values of the frequency $\hat \omega$ lead to consistent evolutions in the gravity side.

For $\hat{\omega} < 1/2$ the WKB expansion given by Eqs.\ (\ref{eq:blinear})-(\ref{eq:slinear}) 
is
 valid in the overlap region since the stationary point lies beyond the horizon,  but expanding both solutions at large $R$ and large $n$ yields inconsistent asymptotics, meaning that no solutions can exist for these values of $\hat{\omega}$. 
 For $\hat{\omega} \ge 1/2$, the WKB solution can be matched to the near horizon dynamics after correctly treating the behaviour of the metric function in the vicinity of the stationary point 
 $r_{\rm S}=2 \hat \omega \Lambda$. A particularly interesting regime is the region $\hat \omega \gtrsim 1/2$, which corresponds to the smallest frequencies where this type of fluctuations occur. In this range the stationary point lies close to the horizon and the near horizon solution Eqs.\ (\ref{eq:ANHnp})-(\ref{eq:B0NRsoln}) 
  must first be matched to the stationary point solution given by Eqs.\ (\ref{eq:blinearIR})-(\ref{eq:slinearIR}), which in turn can be matched to the near horizon solution. 
  In what follows, consistent with our definition of $x$, 
we will expand\footnote{%
Including a $n^{-1/3}$ contribution to $\hat{\omega}$ prevents a consistent matching between the stationary point solution given by Eq.\ (\ref{eq:blinearIR}) and the WKB solution given by Eq.\ (\ref{eq:blinear}). 
} 
\be
\label{eq:hatomegaexp}
\hat{\omega} = \frac{1}{2} + n^{-2/3} \delta \omega_1+ n^{-1} \delta \omega_2 + n^{-4/3} \delta \omega_3 + ... \, ,
\ee 
similarly to the analysis of the QNM frequencies of 
 a static black-brane background in Appendix \ref{app:QNM}.
Note that the fact that $\hat \omega -1/2 \sim \mathcal{O}(n^{-2/3})$ justifies the use of the near boundary approximation, Eqs.\ (\ref{eq:blinearIR})-(\ref{eq:slinearIR}), to describe the dynamics close to the stationary point, since 
\be
\label{eq:AiryMatval}
\left(\frac{r_{\rm S}}{\Lambda}\right)^n \approx \left(1+ \delta \omega_1 \frac{2}{n^{2/3}} \right)^n \underrel{n\to \infty}{=}  e^{2\delta \omega_1 n^{1/3}} \gg1\,,
\ee
where $\delta \omega_1$ is positive, since  $\hat \omega \gtrsim 1/2$.

Examining the stationary point solution given by Eq.\ (\ref{eq:blinearIR}) at large $R$ and large $n$ we find,
\begin{align} \label{IRlimit}
 \lim_{R\rightarrow \infty}&\frac{1}{r^n} b(r,\tau) \propto \, \garb_0(\tau) R^{- \frac{1}{2} + \frac{i}{2}} \\ & \times\left(\text{Ai}\left(-2^{2/3} \delta \omega_1 \right)+\frac{ \text{Ai}\left(-2^{2/3} \delta \omega_1 \right)\left( ... \right) +\frac{\text{Ai}'\left(-2^{2/3} \delta \omega_1\right) \left(-\frac{2 i
   \garb_0'(\tau)}{\garb_0(\tau)}-2 \delta \omega_2 +\log (R)-\frac{i}{1 + \Lambda \tau}\right)}{\sqrt[3]{2}}}{\sqrt[3]{n}} \right) + O(n^{-2/3}) \, . \nonumber
\end{align}

Similarly, near the special value of $\hat{\omega} = \frac{1}{2} + \delta$ with $\delta > 0$, Eq. (\ref{eq:B0NRsoln}) has a large $R$ expansion of the form, 
\begin{equation} \label{NZlimit}
\lim_{R\rightarrow \infty}B_{0 j}(R,\tau) = f(\tau) \frac{\Gamma (1-i)}{\Gamma \left(\frac{1}{2}-\frac{i}{2}\right)^2} \left[(1+ \Lambda \tau) R\right]^{-\frac{1}{2}+\frac{i}{2}} \left(-2 H_{-\frac{1}{2}-\frac{i}{2}} + \log (
   (1+ \Lambda \tau) R)\right) + O(\delta)\, ,
\end{equation}
where $H_{m}$ denotes the $m^{\text{th}}$ Harmonic number. Immediately by comparing the logarithmic structure of Eqs.\ (\ref{IRlimit}) and (\ref{NZlimit}) at fixed orders in $n$, we can set $-2^{2/3}\delta \omega_1$ to be the zeros of the Airy Function and fix the function $\garb_0(\tau)$ through,
\begin{equation}
-\frac{2 i \garb_0'(\tau)}{\garb_0(\tau)}- 2 \delta \omega_2-\frac{i}{1+ \Lambda \tau}= - 2 H_{-\frac{1}{2}-\frac{i}{2}} + \log (1+ \Lambda \tau)  \, ,
\end{equation}
which yields,
\begin{equation}
\garb_0(\tau)=\tilde C_0 (1+ \Lambda \tau)^{-\frac{1}{2}+\frac{i}{2}+\frac{i \Lambda \tau}{2}} \exp{\left(-\frac{i}{2} (1-2 (\delta \omega_2 - H_{- \frac{1}{2} - \frac{i}{2}}))\Lambda \tau \right)}
\end{equation}  
where $\tilde C_0$ is an undetermined overall constant related to initial data.\footnote{Comparing the overall time dependence of Eq.'s (\ref{IRlimit}) and (\ref{NZlimit}) will fix,
\begin{equation}
f(\tau) \propto (1+\Lambda \tau)^{\frac{i \Lambda \tau}{2}} \exp{\left(-\frac{i}{2} (1-2 (\delta \omega_2 - H_{- \frac{1}{2} - \frac{i}{2}})) \Lambda \tau\right)} \, ,
\end{equation}
where the proportionality can be fixed by accounting for an overall constant shift between Eqs.\ (\ref{OmegaAnsatz}) and (\ref{eq:OmegaNH}).}
 Using \eqn{eq:mnbairy} and \eqn{eq:deltaepsilonh}, the energy density of this associated fluctuation in the boundary theory is 
 \be
 \label{eq:deltaepsilonfinal}
 \delta \tilde \epsilon(\tau) = C_0 \Lambda^n \left( \Lambda \tau\right)^{-3/2}   (\Lambda \tau + i)^{\frac{i}{2}(\Lambda \tau +i)}   \exp\left[-i \Lambda\tau \left( \frac{ n+1}{2} + \delta \omega_1 n^{1/3} +H_{- \frac{1}{2} - \frac{i}{2}} \right) \right] \,, 
 \ee
where the constant $C_0$ is related to $\tilde C_0$ by constant factors involving $n$, and $\delta \omega_1$ is given by a zero of the Airy Ai function. 
 
 The inspection of \eqn{eq:deltaepsilonfinal} clearly shows that this type of fluctuation of the energy density  is non-perturbative in the $1/D$ expansion. Indeed, for times parametrically of the same order as $\Lambda$, these fluctuations oscillate very quickly and they become blind to the $1/D$ expansion.
 Note that while the leading contribution to the oscillation frequency 
  in the regime we have considered is fixed,\footnote{%
  Additional oscillation modes can exist with $\hat \omega >1/2$, which we have not analysed.
  }
 $i (n+1) \Lambda/2$, these types of fluctuation possess different oscillation frequencies, distinguishable at subleading order and given by the discrete set of zeros of the Airy Ai function. 
 Therefore, when the fluctuations are small, we can characterise the state by an infinite set of constants, each associated with one of the modes. These sets of constants codify the many possible off-equilibrium initial states that can be prepared for the gauge theory dual at any fixed spacetime dimension.  
 
 It is also instructive to evaluate the $n^0$ contribution to the exponential factor.  While the leading and next to leading oscillation frequencies are purely real, the dissipation of the oscillation only occurs at subleading order, since 
 \be
 H_{- \frac{1}{2} - \frac{i}{2}}  \approx  -0.290892 - 1.44066 i\,,
 \ee
 which, at this order,  is common to all of these fluctuating modes. This uniformity implies that all these fluctuation dissipate at the same time scale,  of order $1/\Lambda$.

 The characteristic oscillation frequencies of these fluctuations coincide, up to some trivial kinematic factors whose origin will become apparent in the next subsection, with the QNM frequencies of a static black-brane computed in \App{app:QNM}. This outcome is expected, since experience with this set-up at finite $D$ indicates that the late-time, non-hydrodynamic evolution is controlled by 
 the static QNMs. 
  At late times, $\tau\gg 1/\Lambda$, \eqn{eq:deltaepsilonfinal}
 may be expressed as a power series in inverse powers of $\tau$, in which the leading order is given by
 \be
 \label{latetimelargenenergydensity}
  \delta \tilde \epsilon(\tau) \underrel{\tau\to \infty}{=} \Lambda^n C_0   
  e^{-i \Lambda\tau \left( \frac{ (n+1)}{2} + \delta \omega_1 n^{1/3} +H_{- \frac{1}{2} - \frac{i}{2}} \right) } 
  \,
  \left ( \Lambda \tau  \right)^{i \Lambda \tau/2}
  \,
 \frac{1}{\left(\Lambda \tau\right)^2 }\,, 
 \ee
 which matches the hydrodynamic analysis we provide in the next subsection.  The subleading terms in $\tau$ correspond to higher order terms in the gradient expansion, and the all orders resummation that the large $D$ calculation provides gives access to the physics at early times.  \green{[[ some changes ]]}

\subsection{Hydrodynamic trans-series at arbitrary fixed $D$} 
\label{subsec:hydrotrans}

 To better understand the origin of the non-perturbative in  $1/D$ contributions to the energy density evolution identified above, in this section 
we compare those results with the expectations from the gradient expansion at fixed $D$. As we have seen in \Sec{sec:LDG}, 
the power expansion in $1/D$ of boost invariant longitudinal expansion is fully controlled by the hydrodynamic limit. Since, as we have argued, the non-pertubative contribution \eqn{eq:deltaepsilonfinal} is sensitive to non-hydrodynamic QNMs, these contributions cannot be captured by a mere gradient expansion. The fact that \eqn{eq:deltaepsilonfinal} in addition to being non-perturbative in the $1/D$  expansion is also non-perturbative in the gradient expansion $u\sim 1/\tau$ makes this observation apparent. Nevertheless, in the gradient analysis, non-perturbative contributions in gradients in the form of a trans-series have been identified \cite{Heller:2015dha,Basar:2015ava,Florkowski:2017olj,Spalinski:2018mqg,Aniceto:2018uik}, 
which control the behaviour of large order gradient perturbations \cite{Heller:2013fn}. In this subsection we explore the connection between these two types of non-perturbative contributions.

We start by performing the gradient trans-series analyses at arbitrary $D$. Following  \cite{Heller:2015dha,Basar:2015ava,Florkowski:2017olj,Spalinski:2018mqg,Aniceto:2018uik},
 at sufficiently late times we can supplement the gradient  expansion of the different metric functions~(\ref{eq:Aexp})-(\ref{eq:dexp}) with non-perturbative (in gradients) contributions in the form of a trans-series
\bea
\bar{A}(r,\tau) & =& \sum_{i=0}^\infty u^i \bar{A}_{i 0}(s) +  \sum_j \Omega^{\rm LT}_j(u) \sum_{i=0}^\infty u^i \bar{A}_{i j}(s)~,\, ...
\label{eq:Aexptrans} \\
\bar{B}(r,\tau) & =& \sum_{i=0}^\infty u^i \bar{b}_{i 0}(s) + \sum_j \Omega^{\rm LT}_j(u) \sum_{i=0}^\infty u^i \bar{b}_{i j}(s)~, \, ...
\label{eq:Bexptrans} \\
\bar{d}(r,\tau) & = &\sum_{i=0}^\infty u^i \bar{d}_{i 0}(s) + \sum_j \Omega^{\rm LT}_j(u)\sum_{i=0}^\infty u^i \bar{d}_{i j}(s)~, \, ...
\label{eq:dexptrans}
\eea
where the ellipses denote higher order terms in the trans-series, which arise as powers of $\Omega^{\rm LT}(u)$, a non-analytic function of $u$ which may be parametrised as 
\bea
\Omega^{\rm LT}_j(u)&\equiv& \exp\left\{-i \int d\tau \bar \omega^{(j)} \tilde \epsilon^{1/n}\right\} 
\nonumber \\
&\approx& u^{-\frac{2 i\bar \omega^{(j)}}{(n-2) n}}  e^{ -i\frac{ \bar \omega^{(j)}}{u} \frac{n-1}{n-2} }
\left(1-i\bar \omega^{(j)} \left( \frac{1}{n^3-3 n^2+2 n}-\frac{\beta
   }{(n-1) n}\right) u+ \mathcal{O} (u^2) \right)\,,
\eea
with $\bar \omega^{(j)}$ a (set of complex) number(s). In the second line we have used the hydrodynamic approximation (\ref{eq:gradexpe}) to approximate the late time energy density. 
As stated, for non-zero $\bar \omega^{(j)}$ this contribution is non-analytic in the normalised gradient $u$.  
 Inserting the above ansatz into Einstein's equations and expanding
to leading order in the normalised gradient $u$ we find that two of the metric functions do not get leading order trans-series corrections 
\be
\label{eq:adleadingtrans}
\bar{A}_{0 j}  = 0 \,, \quad \bar{d}_{0 j}  = 0 \, ,
\ee
while the leading order trans-series correction to $B$ satisfies the equation
\begin{equation}
\label{eqn:QNMlt}
\bar{b}_{0 j}''(s) + \frac{\left((n-1) + s^n - 2 i \bar \omega_{j} \,s \right)}{s(-1 + s^n) } \bar{b}_{0 j}'(s) + \frac{(n-1)i \bar  \omega_{j}}{s(-1 + s^n) } \, \bar{b}_{0 j}(s)  = 0 \, ,
\end{equation}
which coincides with the equation for tensor fluctuations of the black-brane at zero spatial momentum (see Appendix~\ref{app:QNM}). 
Since the metric functions must be normalisable and regular at the horizon, the set $\bar \omega^{(j)}$ is given by the non-hydrodynamic quasinormal modes of the black brane at zero spatial momentum
normalized by the $n$-th root of the equilibrium energy density (\ref{eq:etoTrel}), $\bar \omega^{(j)} = \omega/\tilde \epsilon^{1/n}_{\rm eq}= n \omega/4 \pi T $. 
  Since at subsequent order in the gradient expansion the functions $A_{i j}$ with $i>0$ do not vanish, the structure of \eqn{eq:Aexptrans} and the leading 
 order solution \eqn{eq:adleadingtrans} implies that energy density in the dual gauge theory possesses arbitrary non-perturbative in gradients contributions in the form
 \be
 \label{eq:epsilontrans}
 \tilde \epsilon(u) =  \tilde \epsilon_{\rm hyd} (u) +\Lambda^n \sum_{j} C_j  u^{\frac{2}{n-2}\left( n-1-\frac{ i\bar \omega^{(j)}}{n} \right)}  e^{ -i\frac{ \bar \omega^{(j)}}{u} \frac{n-1}{n-2} } \left(1+\sum_{i=1} \rho_i u^i \right)
 \ee
 with $\epsilon_{\rm hyd} (u)$ the hydrodynamic expansion~(\ref{eq:gradexpe}), and $\rho_i$ is a set of  fixed  coefficients determined by the subsequent orders in the tran-series expansion. The complex numbers $C_j$ are, on the contrary, arbitrary and they depend on the 
   initial conditions of the evolution.

Let us now address the large-$D$ limit of the expression for the energy density~(\ref{eq:epsilontrans}). 
As we have seen in the previous section and in \App{app:QNM}, the QNM spectrum can be computed in series of inverse fractional powers of $n$ \cite{Emparan:2014aba,Emparan:2015rva}. For the least damped modes, $ \bar \omega^{(j)}= \hat \omega \Lambda$ with $\hat \omega$ given in \eqn{eq:hatomegaexp}.
Taking the large-$D$ limit of \eqn{eq:epsilontrans}, the contribution of the least damped modes to the energy density leads to 
\be
\label{eq:epsilontransgradfinal}
\tilde \epsilon(u) = \Lambda^n \left(\frac{1}{\tau \Lambda} + \sum_j C_j e^{- i \left(\frac{n+1}{2} + \delta \omega^{(j)}_{1} n^{1/3}  + \delta \omega_2 \right) \tau \Lambda} \left(\tau \Lambda\right)^{-2+ i\tau\Lambda /2} \left(1+ \sum_{i=1} \frac{\rho
^\infty_i}{\left(\Lambda \tau\right)^i}\right) +... \right) \,,
\ee
where the ellipses denote subleading $n$ terms and $\rho_i^\infty$ are the large-$D$ limits of the constants $\rho_i$. 
At late times, this expression matches exactly the time dependence of the energy density (\ref{latetimelargenenergydensity}) we found in a large $n$ expansion at late times in the previous subsection.  
But going beyond the strict late time limit, the large-$D$ analysis provides a resummation of all the gradient terms at leading order in $n$, since equating \eqn{eq:epsilontransgradfinal} and \eqn{eq:deltaepsilonfinal} yields
\be
\label{eq:infinitygradients}
\left(1+\frac{i}{\Lambda  \tau
   }\right)^{-\frac{1}{2}+\frac{i
   \Lambda  \tau }{2}} \sim \left(1+ \sum_{i=1} \frac{\rho
^\infty_i}{\left(\Lambda \tau\right)^i}\right) \,.
\ee
Note that at this order in the $1/D$ expansion, the resummation of the gradient terms of the trans-series  elements  is the same for all the QNMs in the sector we have considered in this work.
From (\ref{eq:infinitygradients}), it is clear that an infinite number of gradient corrections in $u\sim1/\Lambda \tau$ occur in the leading order in $1/D$ contribution to the trans-series.  
This infinity stands in contrast to the perturbative results in Section \ref{sec:perturbative}, where we saw that at a given order in the $1/D$ expansion, only a finite number of gradient terms contributed.

\section{\label{sec:discussion}Discussion}

As we have seen, the large-$D$ analysis of Bjorken flow is controlled by hydrodynamics. This result is easy to understand and is a consequence of the way we have taken the large-$D$ limit. As we have discussed, to be able to compare holographic theories with different numbers of dimensions, we have demanded that all these theories possess the same late-time, hydrodynamic behaviour, which fixes the typical energy scale $\Lambda$, defined in \eqn{eq:Lambdadef}. This choice forces us to consider times for which the product $w_\epsilon\equiv \tau \tilde{\epsilon}^{1/n}\sim\mathcal{O}(n^0)$. Given the separation between the effective temperature and the $n$-th root of the energy density~(\ref{eq:etoTrel}), the time scale 
$\tau \sim 1/\Lambda$ 
under consideration in this work is parametrically separated from the microscopic scale $\tau_\mu \sim 1/T\sim \tau/n$ when all non-hydrodynamic modes decouple. Since $1/\tau_\mu$ is also the size of the typical gradient, these considerations are similar to other large-$D$ setups \cite{Emparan:2013oza,Romero-Bermudez:2015bma,Emparan:2015rva,Andrade:2015hpa,
Herzog:2016hob,Rozali:2017bll,Emparan:2016sjk,Iizuka:2018zgt,Tanabe:2015hda,Bhattacharyya:2015fdk,
Tanabe:2016opw,Rozali:2016yhw,Dandekar:2016fvw,Dandekar:2016jrp,Bhattacharyya:2016nhn,
Chen:2017wpf,Chen:2017hwm,Bhattacharyya:2017hpj,Miyamoto:2017ozn,Chen:2017rxa,Herzog:2017qwp,
Dandekar:2017aiv,Emparan:2018bmi,Andrade:2018zeb,Andrade:2018nsz}, 
where hydrodynamics also describes the dynamics of the large-$D$ limit. 

In contrast to many of those setups, our leading order dynamics (in the $1/D$ expansion) are  identical to ideal hydrodynamics. This can be easily understood from the $D$-scaling of the different quantities. 
Typical large-$D$ setups (see e.g.\ \cite{Herzog:2016hob,Rozali:2017bll,Emparan:2016sjk,Emparan:2018bmi,Andrade:2018zeb})
focus on slow dynamics,  with energy flux velocities $T^{0i}/\epsilon \sim \mathcal{O}(1/\sqrt{n})$ and small rate of change of energy $\del_t \ln \epsilon \sim \Lambda/n$
 which, together with the transport coefficients~(\ref{eq:tcoefgen}), lead to viscous contributions at leading order in $n$. 
Bjorken dynamics are, on the contrary, fast, since the velocity flux can take any value away from mid-rapidity and the rate of change $\del_\tau \ln \epsilon \sim 1 /\tau \sim \Lambda$ for the 
time scales we consider. 
As discussed in \Sec{sec:HydPrem}, these scalings imply that viscous dynamics appear only at next-to-leading order. At this order, however, second order hydrodynamics captures all the evolution of the system, consistent with the constraints on the transport coefficients imposed by the leading order analysis of slow dynamics at large-$D$.

By extending the large-$D$ expansion of Bjorken flow up to next-to-next-to-next-to-leading order $\mathcal{O}(n^{-3})$ we have managed to extract the large-$D$ behaviour of a certain subset of higher-order transport coefficients up to $6$-th order in gradients. Denoting these coefficients by $\lambda^{(i)}$, with $i$ the gradient order, these are related to the constants $\theta^{(i)}$, defined in \eqn{eq:DeltaPextended}, as 
\be
\lambda^{(i)} = \theta^{i} \tilde{\epsilon}^{1-i/n}_{\rm eq} \,,
\ee
with the equilibrium energy density given in \eqn{eq:etoTrel}. With this normalisation 
$\lambda^{(1)}\equiv -2 \nu \eta$, 
$\lambda^{(2)}\equiv
-2\nu^2 \left(\eta \tau_\pi -2 (n-3)\lambda_1 /(n-2)\right)
$
and $\lambda^{(3)}$ is a combination of 14 third-order coefficients in the basis of \cite{Grozdanov:2015kqa}. Beyond third order the coefficients are not classified. 
Nevertheless, in a basis in which only space-gradients (in the fluid rest frame) of the velocity and energy density field are employed, these $\lambda^{(i)}$ are non-linear combinations of gradients of the velocity field without vorticity contributions. The finite number of inverse powers of $\tau$ in our result \eqn{eq:athirdordern}
also implies that in units of $\tilde {\epsilon}^{1/n}_{\rm eq}$ relevant transport coefficients beyond $6$-th order are suppressed by at least $n^{-4}$. Note  that the leading $D$-dependence of combinations of higher dimensional transport coefficients can also be determined from the $D$-expansion of the dispersion relation of hydrodynamic modes in \cite{Emparan:2015rva}. However, the set of transport coefficients that those relations
 have access to 
 is different from the relevant combinations for Bjorken flow. 
 Since sound waves are linearised perturbations, 
 the relevant corrections are linear combinations of space gradients (in the fluid rest frame) of the velocity field or energy density fluctuations.  In contrast, in homogeneous Bjorken flow 
 space gradients of the energy density vanish and the relevant corrections are non-linear combinations of the velocity field. 
Therefore, this boost invariant analysis constrains a complementary set of transport coefficients.

While the procedure we have followed could in principle be continued for arbitrarily large orders in the $D$-expansion, this boost invariant set-up clearly shows that the $1/D$ expansion does not converge, as already noted in \cite{Emparan:2015rva}. 
In \Sec{sec:NHM}, we have identified a set of boost-invariant excitations that are non-perturbative in the $1/D$ expansion.  Those excitations correspond to 
fast modes, with a typical rate of variation of order $n \Lambda$, parametrically separated from the hydrodynamic regime. As we have checked explicitly, those 
correspond to non-hydrodynamic excitations, governed by the large-$D$ limit of quasi-normal modes of black-branes. In addition to being non-perturbative in the $1/D$
expansion, those contributions are also non-perturbative in the gradient expansion and at fixed $D$ they are responsible for the non-convergence of the hydrodynamic 
series \cite{Heller:2015dha,Basar:2015ava,Florkowski:2017olj,Spalinski:2018mqg,Aniceto:2018uik}.  Analogously, 
the $D$-dependence of this non-perturbative component also indicates that studying large orders in the large-$D$ expansion would lead to factorial growth of the $D$-expansion 
coefficients of the energy and, consequently, its Borel transform would have a finite radius of convergence. 
Nevertheless, the non-analytic $D$-dependence of the lowest QNMs may lead to interesting patterns in the behaviour of large order perturbation theory, 
which could be studied after automating the large-$D$ computation in this set-up, as was done for the gradient expansion. 

From the point of view of  time-evolution, the existence of this set of excitations codifies the many possible initial conditions that can be specified for the system. 
The behaviour of the relevant QNMs at large-$D$ also has interesting consequences for the way in which matter becomes independent of the details of the initial conditions at sufficiently late times. Unlike the familiar example of $D=5$ ($n=4$) where the real and imaginary parts of those QNMs are comparable, in the large-$D$ limit these two contributions have parametrically different behaviours. As discussed in \Sec{sec:NHM}, the leading contribution of the least damped modes becomes proportional to the number of space-time dimensions in the dual theory, while the imaginary part saturates into a constant of order $\epsilon^{1/n}$. What this means is that, at the time scale at which we have performed the analysis, the amplitude of the non-hydrodynamic excitations are not suppressed. The reason those modes decouple from the expansion at this time-scale is that they oscillate rapidly about the hydrodynamised evolution. As such, they average to zero over the typical expansion time-scale.
However, the damping rate of those excitations remains of order $\Lambda$, which implies that the amplitude of oscillations is not parametrically suppressed over the time-scales we study. 
If there is a hydrodynamic attractor at large-$D$, this analysis indicates that the approach of specific initial conditions to this attractor is  different from the cases already studied in the literature up to now (see \cite{Florkowski:2017olj} for a review).
It would be interesting to study whether by using $D$ as an expansion parameter, holographic computations in arbitrary space-time dimension could be used to further constrain the dynamics of hydrodynamic attractors.

\acknowledgments

We thank T. Andrade, R. Emparan, N. I. Gushterov, and M. Heller for useful discussions. 
JCS is a University Research Fellow of the Royal Society (on leave) and is supported by  grants SGR-2017-754, FPA2016-76005-C2-1-P and MDM-2014-0367. BM is a Commonwealth Scholar and is also supported by the Oppenheimer Fund Scholarship. CPH is supported in part by the U.S.\ National Science Foundation Grant PHY-1620628 and 
by the U.K.\ Science \& Technology Facilities Council Grant ST/P000258/1.

% -----------------------------------------------------------------------------------------------------
% -----------------------------------------------------------------------------------------------------
% -----------------------------------------------------------------------------------------------------

% -----------------------------------------------------------------------------------------------------
% -----------------------------------------------------------------------------------------------------
% -----------------------------------------------------------------------------------------------------

\appendix
\newpage
\section{Non-Perturbative Solutions in the Near Boundary Region}\label{generalOmegaAnsatz}
In this appendix we justify the form of Eq. (\ref{OmegaAnsatz}) in Subsection \ref{subsec:transNBR}. In the near boundary region, after introducing variables $a$, $b$ and $s$ in Eqs.\ (\ref{eq:linearA}) to (\ref{eq:linearS}), we find Eq.\ (\ref{eq:CY3}) to take the form,
\begin{align} 
\label{eq:CY3linear}
&
 \left( n+1 + \frac{1}{r \tau + 1} \right) \frac{b^{(0,1)}(r,\tau)}{2r}
   +\frac{(n-1) r}{2}
   b^{(1,0)}(r,\tau )- \frac{r^2}{2} b^{(2,0)}(r,\tau
   )-b^{(1,1)}(r,\tau ) 
   \nonumber \\
 & +\frac{1}{r \tau + 1} \frac{n-2}{n-1} \left[  r a^{(1,0)}(r,\tau )- a(r,\tau )+\frac{(n-1) s^{(0,1)}(r,\tau
   )}{r^2 } \right]
  = 0 \, . 
\end{align}
We will look for solutions of the form,
\begin{align} \label{eq:nonPertAnsatzAppendix}
a(r,\tau) & = \Omega(r,\tau) \sum_{i=0} n^{-i} a_{i}(r,\tau) \, , \\
b(r,\tau) & = \Omega(r,\tau) \sum_{i=0} n^{-i} b_{i}(r,\tau) \, , \\
s(r,\tau) & = \Omega(r,\tau) \sum_{i=0} n^{-i} s_{i}(r,\tau) \, ,
\end{align}
where the function $\Omega(r,\tau)$ satisfies,
\begin{equation}
 \frac{\partial_{\tau} \Omega(r,\tau)}{\Omega(r,\tau)} \sim  \frac{\partial_{r} \Omega(r,\tau)}{\Omega(r,\tau)} \sim O(n) \, ,
\end{equation}
and derivatives of all $a_i$, $b_i$ and $s_i$ go like $O(n^{0})$. Boundary energy-momentum conservation imposes that $\del_\tau a= b$ as $r\rightarrow \infty$, so that the leading power of $a_i$ in Eq.\ (\ref{eq:nonPertAnsatzAppendix}) will be suppressed with respect to $b(r,\tau)$. Similarly, imposing the other Einstein's equations implies that the leading order term in $s$ is suppressed by $n^2$. 
Whenever we take a derivative with our Ansatz we pick up an extra factor of $n$ so that only terms proportional to $b$ will remain in the equation of motion at leading order in $n$. 
We can extract $\Omega$ from the leading order in $n$ terms of Eq.\ (\ref{eq:CY3linear}): 
\begin{equation}
\frac{n}{ r}b^{(0,1)}(r,\tau ) +nr
    b^{(1,0)}(r,\tau )-r^2 
   b^{(2,0)}(r,\tau )-2b^{(1,1)}(r,\tau ) = 0 \, . 
\end{equation}
This equation is linear with $\tau$ independent coefficients.
Therefore, we can write all solutions of $\Omega$ in terms of a Fourier transform. Factoring out the leading $n$ coefficient of $\tau$ in this Fourier transform, we can identify,
\begin{equation}
\Omega(r,\tau) = e^{- i n \hat{\omega}  \tau + n f(r)} \, .
\end{equation}

\newpage
\section{\label{app:QNM}Tensor Quasi-Normal Modes of Black-Branes at Large-$D$} 
In this section we describe the quasi-normal spectrum of black-branes in the large-$D$ limit at zero spatial momentum. These modes control the non-perturbative contributions in $1/D$ to the Bjorken flow, as discussed in \Sec{sec:NHM}. As is well known \cite{Kovtun:2005ev} this channel does not possess hydrodynamic excitations and the frequency of all QNMs is finite in the zero momentum limit. 
The equation for these modes can be easily obtained by studying small anisotropic fluctuations of the black-brane. In standard Eddington-Finkelstein form (which differs from the gauge in \Sec{sec:LDG}), to leading order in the perturbation, the metric is
\be
ds^2_{\rm{ B-B+fluc}} = -r^2 A_{\rm BB} (r) d t^2 + 2\, dt dr + r^2 \left(1- \epsilon\,  Z(t,r)  \right) d x_\parallel^2 + r^2 \left(1+\frac{\epsilon}{n-2}   Z(t,r)\right) d x_\perp^2
\ee
with $A_{\rm BB}=1-(\Lambda/r)^n$ the blackening factor of the brane in $D=n+1$ dimension and $\epsilon$ is a book-keeping parameter that indicates that the anisotropic fluctuation $Z(t,r)$ is small. Linearising Einstein's equations in $\epsilon$ and after a Fourier transform  the equation of motion for the fluctuation in terms of the coordinate $z=\Lambda/r$ is given by
\be
 \label{eq:Zgen}
      Z''(z) - \frac{\left((n-1) + z^n - 2 i \bar \omega_{j} \,z \right)}{z(1 - z^n) } Z'(z) - \frac{(n-1)i \bar  \omega_{j}}{z(1 - z^n) } \, Z(z)  = 0 \, ,
\ee
which coincides with the \eqn{eqn:QNMlt} for the leading order trans-series term in Bjorken flow.\footnote{%
The Fourier transform has a plane wave factor of the form $e^{- i \bar \omega_j  \Lambda t}$ such that 
$\bar \omega_j$ is dimensionless. \green{ [[ added ]]}
}
The large-$D$ analysis of this equation is very similar to that of the QNMs of black-holes in 
 \cite{Emparan:2014aba},  which we follow closely.  As in that case, this equation exhibits two distinct regions in the coordinate $z$: the near boundary region $z\rightarrow 0$, where the quickly changing function $z^n$ can be neglected and the dynamics are those of vacuum $AdS_D$; and the near horizon region $z\sim1$ where the dynamics of the fluctuation is sensitive to the black-brane. 

To zoom into the near horizon region, we introduce the variable $R= 1/z^n$, as was done in \Sec{sec:LDG}. 
To find a non-trivial QNM spectrum at large-$D$, we also introduce the scaled variables $\hat \omega= \omega/n$. Taking 
  $n\rightarrow \infty$,  \eqn{eq:Zgen} becomes   
   \be
 \label{eq:Znear}
 \frac{\left(1+2\left( i \hat \omega -1 \right) R \right)  Z_{\text{NH}}'(R)}
 {R \left(1- R \right)}
   -\frac{i \hat \omega Z_{\text{NH}}(R)}{(R -1) R }+Z_{\text{NH}}''(R)=0\,,
 \ee
 which coincides with the equation for the non-perturbative phase, \eqn{eq:phase}. Regular or ingoing 
 solutions of this equation at the horizon, $R=1$, are given by
\be
\label{eq:ZNH}
Z_{\text{NH}}^{QNM}(R)=
\, _2F_1\left(-\frac{1}{2} \sqrt{1-4 \hat \omega^2}-i \hat \omega+\frac{1}{2},\frac{1}{2} \sqrt{1-4 \hat \omega^2}-i \hat \omega+\frac{1}{2};1-2 i \hat \omega;1-R \right)\,,
\ee
with ${}_2F_1$ the ordinary hypergeometric function.
Note that in this limit and to leading order in $1/n$, this near horizon fluctuation does not depend on~$n$.

To obtain the spectrum of QNMs, we must impose normalizability at the boundary. As already mentioned, in this region the dynamics of the fluctuation are  vacuum-like and the effects of the black-brane can be neglected. 
The effects of the black-brane are signaled by the presence of an $z^n$ term in \eqn{eq:Zgen}, which vanishes in the near boundary limit, leading to the simpler equation
\be
 \label{eq:ZgenNB}
      Z''_{\text{NB}}(z) - \frac{\left(n-1  - 2 i \bar \omega_{j} \,z \right)}{z} Z'_{\text{NB}}(z) - \frac{(n-1)i \bar  \omega_{j}}{z} \, Z_{\text{NB}}(z)  = 0 \, .
\ee
Analytic solutions of this equation, in terms of the Bessel-J function can be found
and
analysed as in \cite{Emparan:2014aba}. Here, however, we will perform an equivalent WKB interpretation, 
which can be generalised to the analysis of fluctuations on top of Bjorken flow.
 Following this strategy,
we search for solutions of \eqn{eq:Zgen} in the near boundary region in the form 
\be
\label{eq:Zwkb}
Z_{\text{NB}}^{QNM}(z)=z^{n/2} e^{-i n \hat \omega z} e^{n \sigma (z)} \,. 
\ee
(While it is true that $e^{n \sigma} =  J_{\frac{n}{2}}( n \hat  \omega z)$, for most of the analysis, it will be more convenient to use the WKB form of the wave function).
At large-$D$, we expand $\sigma$ in inverse powers of $n$
\be
\label{eq:sigmaexpansion}
\sigma=\sigma_0(z) + \frac{1}{n} \sigma_1(z) + ... \,,
\ee
where the ellipses denote additional orders in the large $D$-expansion. The equation for $\sigma_0$ is non-linear and is given by
\be
\sigma'_0(z)=\pm\frac{1}{2}\sqrt{\frac{1}{z^2} -4 \hat \omega^2} \ . 
\ee
Only the $``+"$ branch leads to a normalisable solution at the boundary. Note that in this solution there is a special point $z_{\rm T}=1/2\hat \omega$ around which 
the phase varies slowly. This means that the WKB approximation is not valid in the vicinity of this point and may be viewed as the analogue of a stationary point in quantum mechanics. 
There are three distinct regimes which can be analyzed: $\hat \omega < 1/2$, $\hat \omega \gtrsim 1/2$, and $\hat \omega \gg 1/2$.  
For $\hat \omega<1/2$, the stationary point lies in the region where the effect of the
$z^n$ gravitational potential of 
the black-brane
is non-negligible and the WKB approximation is valid in all the near boundary region. However, for this range of values of $\hat \omega$ the solution 
cannot be matched to the near horizon region and no QNMs appear. 
On the contrary, for $\hat \omega \geq 1/2$ the WKB approximation fails
while the black-brane gravitational field is negligible, and QNM frequencies can be obtained by matching the near horizon solutions to the near boundary ones.

The limit $\hat \omega \gg 1$ was analyzed previously by ref.\ \cite{Natario:2004jd}. 
 As the modes are highly damped, they are perhaps physically less interesting.  Additionally, we are specifically interested in the case where $n$, not $\hat \omega$, is the largest parameter in the problem.
   As the calculation is more straightforward than the more interesting case $\hat \omega \gtrsim 1/2$, let us nevertheless summarize the results.
 The near horizon solution (\ref{eq:ZNH}) has a large $\xi$ expansion of the form
 \be
Z_{\text{NH}}^{QNM} \sim R^{-1/2} (-2 i e^{-2 \pi \hat \omega} R^{2 i \hat \omega} + 1 ) \ .
 \ee
 Using the Bessel function form, the near boundary solution has a similar large $R$ expansion
 \be
Z_{\text{NB}}^{QNM} \sim R^{-1/2} (i e^{i n \pi/2 - 2 i n \hat \omega} R^{2 i \hat \omega} + 1) \ .
 \ee
 These two expansions are compatible provided $\hat \omega$ is quantized according to
 \be \label{eq:omegaNS}
 \hat \omega_j = \frac{1}{2} \frac{1}{n + i \pi} \left( 2 \pi j + \frac{n-2}{2} \pi + i \log (2) \right)  \ ,
 \ee
 with $j$ an integer such that $|j| \gg 1$.  
 One aspect of this spectrum that will remain true in the regime $\hat \omega \gtrsim 1/2$ is that the imaginary part is $O(1/n)$ compared with the real part of $\hat \omega.$
 
 The most interesting case\footnote{
 The spectrum  of QNMs with $\hat \omega \gg 1/2$ can be inferred from the analysis of Chamblin-Reall holography performed in \cite{Betzios:2018kwn}, which connects the 
  $\hat \omega \gtrsim 1/2$ region with the asymptotic large frequency region studied in \cite{Natario:2004jd} and summarised in \eqn{eq:omegaNS}.
 } is  $\hat \omega \gtrsim 1/2$. In this case, the WKB expression is valid in the 
entire near boundary region and it only fails in the vicinity of the horizon, $z_{\rm T}\sim1$. 
This range of frequencies also corresponds to the least damped QNMs of the black-brane \cite{Emparan:2014aba}. Using standard arguments, the extent of this region may be determined by 
requiring  $n (\sigma'_0(z))^2\ll \sigma''_0(z)$  which implies that in the region 
\be
\label{eq:WKBinvalid}
 \left | x \right| \ll 1\quad {\rm with } \quad x \equiv \left(\frac{1}{z}-2\hat \omega \right)n^{2/3} \,, 
\ee
the equation for the fluctuations must be solved in a different manner.  Note, in particular, that keeping $x$-fixed while taking the large $n$-limit brings the validity of the WKB region closer and closer to $z_{\rm T}=1$. Therefore, in the region given by \eqn{eq:WKBinvalid} we perform the large-$D$ expansion  with $x$ fixed as 
\be
\tilde {Z}_{\text{NB}}=\sum_{j=0} \psi_j(x) n^{-j/3} \,, 
\ee
where $\tilde {Z}_{\text{NB}}\equiv e^{n \sigma}$. In this expansion, the origin of the fractional powers arises from the definition of $x$ in \eqn{eq:WKBinvalid}. Similarly, we expand the QNM frequency in inverse fractional powers of the number of space-time dimensions
\be
\label{eq:QNMexp}
\omega = n \left(\frac{1}{2} + \delta \omega_1 \frac{1}{n^{2/3}}  + \delta \omega_2 \frac{1}{n} + \delta \omega_3 \frac{1}{n^{4/3}} + \delta \omega_4 \frac{1}{n^{5/3}} + ... \right)\,,
\ee
where the absence of an $n^{-1/3}$ contribution is a consequence of matching this result with the near horizon region \cite{Emparan:2014aba}.  
The validity of the near boundary approximation, \eqn{eq:ZgenNB}, follows from the same argument as in \eqn{eq:AiryMatval}.

Up to order $1/n$, solutions for the functions $\psi_i(x)$ can be found in terms of Airy functions. Imposing that in the $x\gg 1$ region those solutions can be matched to a normalisable mode, expressed in the WKB form of  \eqn{eq:Zwkb} with the expansion \eqn{eq:sigmaexpansion},  and up to an overall normalisation, these functions are
\bea
\label{eq:psi0}
\psi_0(x)&=&\text{Ai}\left(\frac{x}{\sqrt[3]{2}}\right) \,, 
\\
\label{eq:psi1}
\psi_1(x)&=&c_1 \text{Ai}\left(\frac{x}{\sqrt[3]{2}}\right) \,, 
\\
\label{eq:psi2}
\psi_2(x)&=&\frac{1}{20} \left(4
   \text{Ai}\left(\frac{x}{\sqrt[3]{2}}\right) \left(5 c_2+5
   \delta \omega _1+x\right)-2^{2/3} x \left(20 \delta \omega
   _1+7 x\right)
   \text{Ai}'\left(\frac{x}{\sqrt[3]{2}}\right)\right) \,,
   \\
   \label{eq:psi3}
   \psi_3(x)&=&\frac{1}{20} \left(4
   \text{Ai}\left(\frac{x}{\sqrt[3]{2}}\right) \left(5 c_1
   \delta \omega _1+c_1 x+5 c_3+5 \delta \omega
   _2\right) \right.
   \\
   &&
  \hspace{3.5cm} \left. 
   -2^{2/3} x
   \text{Ai}'\left(\frac{x}{\sqrt[3]{2}}\right) \left(20 c_1
   \delta \omega _1+7 c_1 x+20 \delta \omega _2\right)\right) \,,\nonumber
\eea
with $\text{Ai}(x)$ the Airy function of the first kind and $c_1$, $c_2$ and $c_3$ arbitrary constants that need to be determined.
Since the set of Eqs.\ (\ref{eq:psi0})-(\ref{eq:psi3}) extend to the vicinity of the horizon,
we can fix those constants as well as the different QNM frequency corrections $\delta \omega_i$ by matching the linearised wave functions 
to  the near horizon dynamics in the intermediate region, when the coordinate $R$ becomes large and the effect of the black-brane is small. To compare the behaviour close to the stationary point with \eqn{eq:ZNH}, we express Eqs.\ (\ref{eq:psi0})-(\ref{eq:psi3}) in terms of $R$ and
expand in inverse powers of $n$. Note that at leading order, this procedure implies 
\be
\label{eq:xdef}
x=-2 \delta \omega_1 + \mathcal{O}\left(\frac{1}{n^{1/3}}\right)\,. 
\ee
Using Eqs.\ (\ref{eq:Zwkb}), (\ref{eq:psi0}), and (\ref{eq:xdef}),
the leading order in $1/n$  expression for the QNM profile in the intermediate region is given by 
\bea
\label{eq:ZIZfarLO}
Z_{\text{NB}}^{QNM}&=&e^{-i \delta \omega _2-i \delta \omega _1
   n^{1/3}-\frac{i n}{2}-\frac{1}{2}}  
   R ^{-\frac{1}{2}+\frac{i}{2}} 
   \left(\text{Ai}\left(-2^{2/3} \delta \omega
   _1\right)
   + \mathcal{O}\left(n^{-1/3}\right) \,,
\right)
   \nonumber
\eea

We now compare the behaviour of the linearised solution in the intermediate region with that of the near horizon expression, \eqn{eq:ZNH}. The asymptotic behaviour of 
this profile in the limit $R\rightarrow \infty$ is 
\bea
Z_{\text{NH}}^{QNM}&=&
\frac{R^{-\frac{1}{2} \sqrt{1-4 \hat \omega ^2}+i \hat \omega
   -\frac{1}{2}} \Gamma (1-2 i \hat \omega ) \Gamma
   \left(-\sqrt{1-4 \hat \omega ^2}\right)}{\Gamma
   \left(\frac{1}{2} \left(-2 i \hat \omega -\sqrt{1-4 \hat \omega
   ^2}+1\right)\right)^2}+
   \\&&
   \frac{R ^{\frac{1}{2} \sqrt{1-4
   \hat \omega ^2}+i \hat \omega -\frac{1}{2}} \Gamma (1-2 i \hat \omega )
   \Gamma \left(\sqrt{1-4 \hat \omega ^2}\right)}{\Gamma
   \left(\frac{1}{2} \left(-2 i \hat \omega +\sqrt{1-4 \hat \omega
   ^2}+1\right)\right)^2}  \,, 
   \nonumber
\eea
Expanding this expression in the vicinity of $\hat \omega= 1/2 + \delta \omega$,  with $\delta \omega$ small and given by the expansion  \eqn{eq:QNMexp}, the near horizon profile becomes
\bea
\label{eq:ZIZnear}
Z_{\text{NH}}^{QNM} &=& R ^{-\frac{1}{2}+\frac{i}{2}} \left(\frac{\Gamma (1-i) 
   \left(\log (R )-2
   H_{-\frac{1}{2}-\frac{i}{2}}\right)}{\Gamma
   \left(\frac{1}{2}-\frac{i}{2}\right)^2}
\right.
\\
&&
\left. 
   +2 \delta \omega
   \frac{  \Gamma (1-i) }{6 \Gamma
   \left(\frac{1}{2}-\frac{i}{2}\right)^2} 
   \left(c_{\rm L\, 0} +  c_{\rm L\, 1} \log R + c_{\rm L\, 2} \log R^2 -\frac{1}{2}\log R^3   \right) \,, \nonumber
   \right)
\eea
with
\bea
c_{\rm L\, 0} &=&2 H_{-\frac{1}{2}-\frac{i}{2}} \left(2
   \left(H_{-\frac{1}{2}-\frac{i}{2}}\right){}^2+\pi ^2-6 i
   \psi ^{(0)}\left(\frac{1}{2}-\frac{i}{2}\right)+6 i \psi
   ^{(0)}(1-i)-3 \psi
   ^{(1)}\left(\frac{1}{2}-\frac{i}{2}\right)\right) \nonumber
\\
&& 
   +8 \zeta
   (3)+6 i \psi
   ^{(1)}\left(\frac{1}{2}-\frac{i}{2}\right)+\psi
   ^{(2)}\left(\frac{1}{2}-\frac{i}{2}\right) \,,
   \\
c_{\rm L\, 1} &=& -6 \left(H_{-\frac{1}{2}-\frac{i}{2}}\right){}^2-6 i \gamma_E
   -\pi ^2-6 i \psi ^{(0)}(1-i)+3 \psi
   ^{(1)}\left(\frac{1}{2}-\frac{i}{2}\right) \,,
   \\
c_{\rm L\, 2} &=&   
   3 H_{-\frac{1}{2}-\frac{i}{2}}+3 i \,, 
\eea
where $H_n$ is the $n^{th}$-Harmonic number, $\gamma_E$ is the Euler-Mascheroni constant, $\zeta(s)$ is the Riemann~$\zeta$-function  and $\psi^{(i)}$ is the polygamma function of order $i$.

Comparing the expected behaviour from the near horizon, \eqn{eq:ZIZnear}, and near boundary, \eqn{eq:ZIZfarLO}, analyses, in the intermediate region it becomes apparent that these two 
functional forms can only be made compatible provided the leading order term in \eqn{eq:ZIZfarLO} vanishes, which implies that $-2^{2/3}\delta \omega_1$ is a zero of the Airy function, which may be approximated by \cite{Emparan:2014aba}
\be
\label{eq:do1}
\delta \omega_1 \simeq  \left(\frac{3\pi}{16} \left(4 k -1\right)\right)^{2/3} \, , \quad k=1,\, 2, \, .... \,,
\ee
which becomes a more accurate approximation as  $k$ grows.

Having fixed the leading order correction to the QNM frequency, the functional forms of both expressions \eqn{eq:ZIZnear} and \eqn{eq:ZIZfarLO} can be matched order by order in  the $1/n^{1/3}$-expansion. This procedure fixes the corrections of the QNM frequencies
\bea
\label{eq:do2}
\delta \omega_2&=&H_{-\frac{1}{2}-\frac{i}{2}}\,,
\\
\label{eq:do3}
\delta \omega_3&=&\frac{3 \delta\omega_1^2}{5} \,,
\\
\label{eq:do4}
\delta \omega_4&=&\frac{1}{6}\delta \omega_1\left(12
   H_{-\frac{1}{2}-\frac{i}{2}}-8 \zeta (3)-6 i \psi
   ^{(1)}\left(\frac{1}{2}-\frac{i}{2}\right)-\psi
   ^{(2)}\left(\frac{1}{2}-\frac{i}{2}\right)\right) \,, 
\eea
as well as the undetermined constants in \eqn{eq:psi1} and \eqn{eq:psi2}
\bea
c_1&=&\frac{3 i \delta \omega_1^2}{5} \,, \\
c_2&=&
-\frac{1}{150} \delta \omega _1 \left(27 \delta \omega _1^3
   +50 \pi^2-300 i \gamma_E
   \right .
   \\
   &&
   \left.
\hspace{1cm}  +5
   \left(40 i \zeta (3)+42-120 i \psi
   ^{(0)}\left(\frac{1}{2}-\frac{i}{2}\right)+60 i \psi
   ^{(0)}(1-i)
   \right. \right.  \nonumber
   \\
   &&
   \left. \left.
   \hspace{4cm}- 60 \psi
   ^{(1)}\left(\frac{1}{2}-\frac{i}{2}\right)+5 i \psi
   ^{(2)}\left(\frac{1}{2}-\frac{i}{2}\right)\right)
   \right) \,. \nonumber
\eea
Beyond the stated order, the calculation becomes sensitive to both $1/n$ corrections to the near horizon wave function and to additional powers in the expansion of the QNM frequency.

The results for $\delta \omega_1$ and $\delta \omega_2$ are the same that we found in the computation of the trans-series for Bjorken flow in Section \ref{sec:NHM}.  Here, however, due to the simplicity of the differential equation (\ref{eq:Zgen}), we were able to go to higher order and recover $\delta \omega_3$ and $\delta \omega_4$ as well.  
To gain confidence in the results, in \fig{fig:ldm} we compare with numerical evaluations of the QNM spectrum for black-branes for  $n=4$ to 300.  The match is quite good for the low-lying modes, and becomes better with increasing $n$.

\begin{figure}[t]
\begin{center}
\includegraphics[width=.45\textwidth]{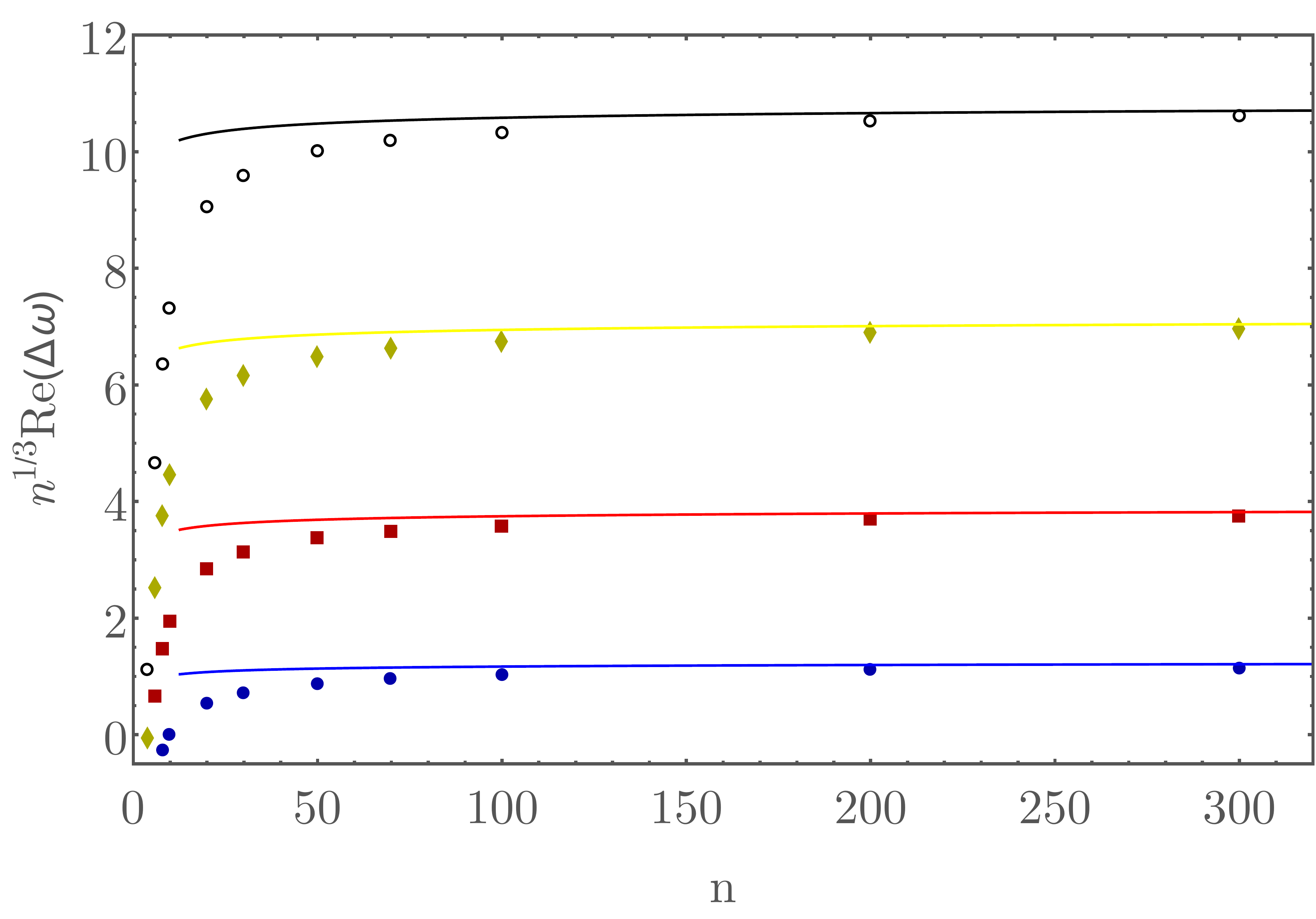}
\includegraphics[width=.45\textwidth]{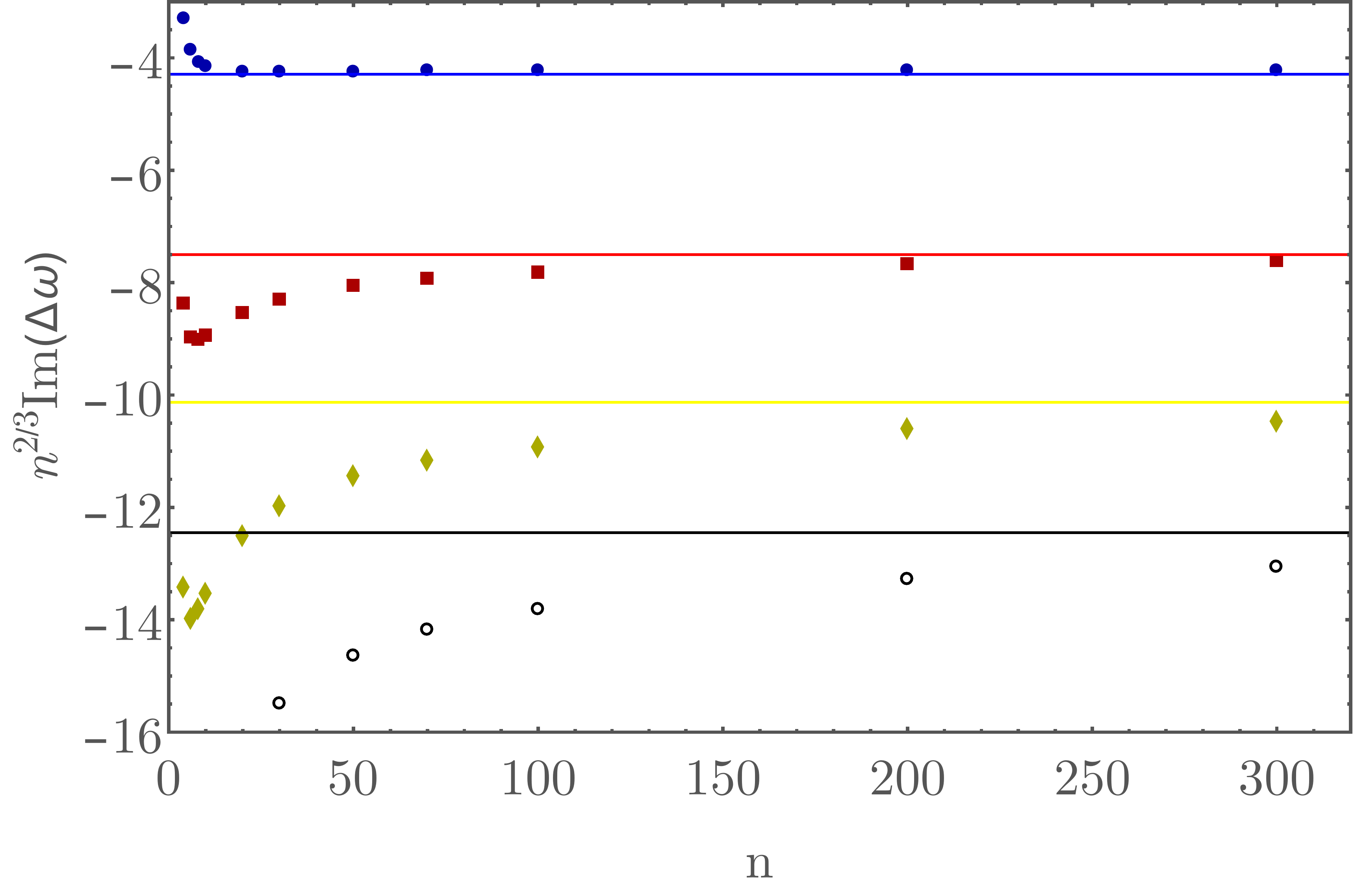}
\caption{\label{fig:ldm} 
Real 
(left) and imaginary (right) parts of the first four quasi normal modes of  black-branes in different numbers of dimension $D=1+n$ in the spin-2 channel at zero spatial momentum. 
In both panels
$\Delta \omega= \omega_{\rm QNM} - n \left(1/2 + \delta \omega_1 n^{-2/3}   + \delta \omega_2  n^{-1} \right) $, with $\delta \omega_1$ the corresponding zero of the Airy function and $\delta \omega_2$ given in \eqn{eq:do2}. The lines correspond to the real and imaginary parts
of $\Delta \omega_{\rm LD}\equiv  n \left( \delta \omega_3 n^{-4/3} + \delta \omega_4 n^{-5/3} \right)$ with the expression for each coefficient given in Eqs.\ (\ref{eq:do3}) and (\ref{eq:do4}) multiplied by the power of $n$ stated in the $y$-axis.
}
\end{center}
\end{figure}
\begin{figure}[t]
\begin{center}
\includegraphics[width=.45\textwidth]{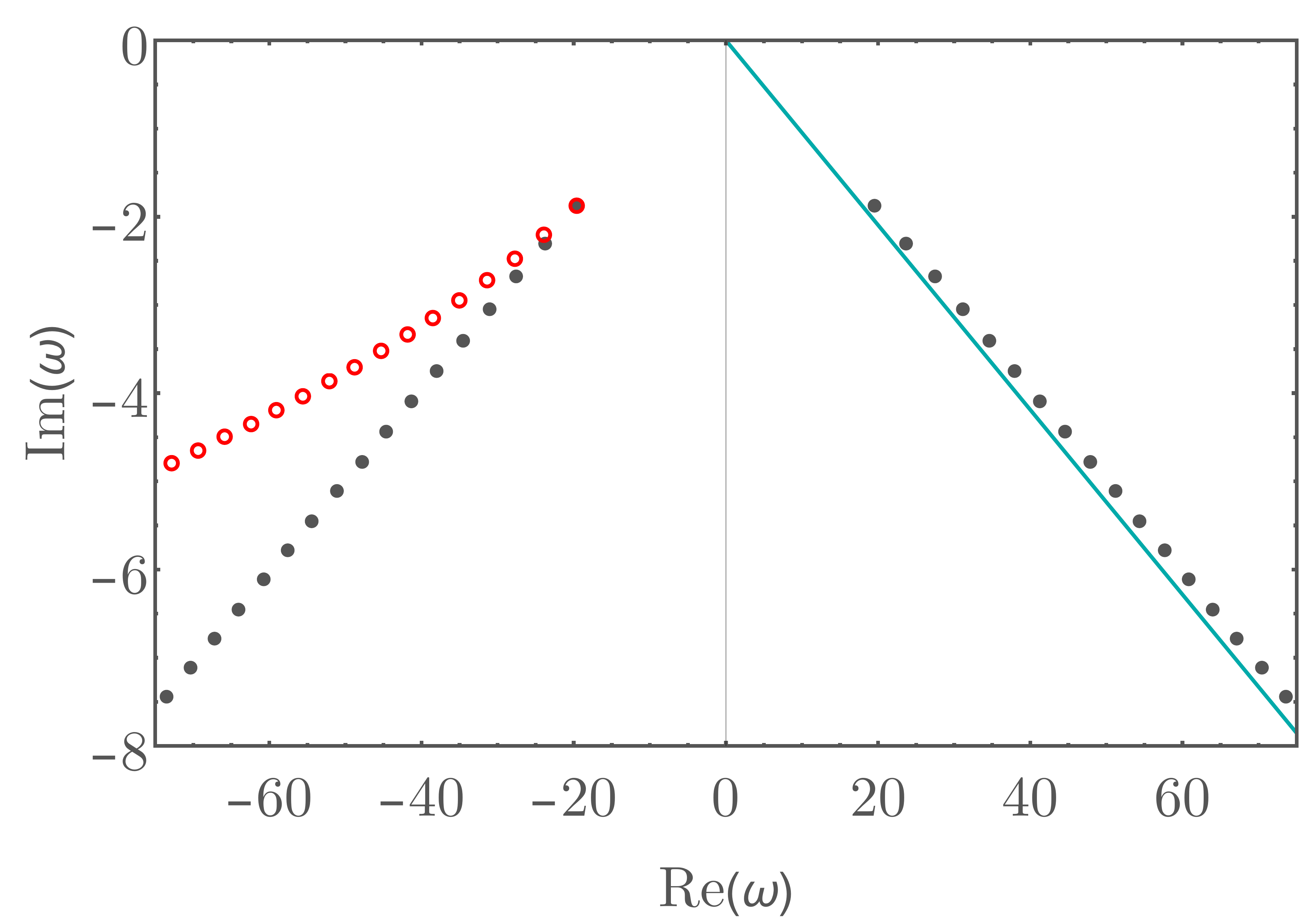}
\caption{\label{fig:ldm} 
The quasinormal mode spectrum for $n=30$.  The solid dots are numerically determined.
For the modes with negative real part, the open circles indicate the prediction from our WKB analysis for modes with $|\omega| \gtrsim n/2$.   
For the modes with positive real part, we have plotted the line $\operatorname{Im}(\omega) = -\frac{\pi}{n} \operatorname{Re}(\omega)$, determined from the $|\omega| \gg n/2$ analysis of the quasinormal mode spectrum.
}
\end{center}
\end{figure}

\section{\label{app:longexpressions}Expressions for the Third-order Expansion of the Different Metric Functions}
In this appendix we tabulate the third-order in the large-$D$ expansion for the different metric functions, found in \Subsec{subsec:NHR}. 

\bea
 A_3&=&  - \frac{\anhth(\tau)}{R} \nonumber \\
 &&\hspace{-0.5cm} +
\frac{4
   \Lambda  \tau  \text{Li}_3\left(\frac{1}{R
   (\Lambda  \tau +1)}\right)}{R (\Lambda  \tau
   +1)^4}+\frac{2 \Lambda  \tau 
   \text{Li}_2\left(\frac{1}{R (\Lambda  \tau
   +1)}\right) \left(2 \left(1-\frac{1}{R
   (\Lambda  \tau +1)}\right)-\frac{\log
   \left(\frac{1}{R (\Lambda  \tau +1)}\right)}{R
   (\Lambda  \tau +1)}\right)}{(\Lambda  \tau
   +1)^3} \, 
   \nonumber 
   \\
   &&\hspace{-0.5cm} -\frac{\left(\Lambda ^2 \tau ^2-4
   \Lambda  \tau +1\right) \log ^3(R)}{6 R
   (\Lambda  \tau +1)^4}-\frac{\log ^2(R)
   \left(\left(\Lambda ^2 \tau ^2-1\right) \log
   (\Lambda  \tau +1)+\Lambda  \tau  (\Lambda 
   \tau -5)\right)}{2 R (\Lambda  \tau
   +1)^4}\, 
   \nonumber\\
   &&\hspace{-0.5cm} -\frac{\log (R) \left(-2 \Lambda  \tau 
   (\Lambda  \tau +2)+(\Lambda  \tau +1)^2 \log
   ^2(\Lambda  \tau +1)-2 (5 \Lambda  \tau +2)
   \log (\Lambda  \tau +1)\right)}{2 R (\Lambda 
   \tau +1)^4} \, 
   \nonumber
   \\
   &&\hspace{-0.5cm} + \frac{4 \Lambda  \tau 
   \left(1-\frac{1}{R (\Lambda  \tau +1)}\right)
   \log \left(\frac{1}{R (\Lambda  \tau
   +1)}\right) \log \left(1-\frac{1}{R (\Lambda 
   \tau +1)}\right)}{(\Lambda  \tau +1)^3} \,, 
\eea
with $\anhth(\tau)$ given in \eqn{eq:a3nh} and the metric function $B_3$ is
\bea
B_{3}& = & -\frac{2 (\Lambda  \tau -1)^2 \text{Li}_3\left(\frac{1}{\Lambda  \tau  R+R}\right)}{(\Lambda  \tau +1)^3}+\frac{4 (\Lambda 
   \tau -1) \text{Li}_3(1-R (\Lambda  \tau +1))}{(\Lambda  \tau +1)^3}  \\
   && -\frac{2 \Lambda  \tau 
   \text{Li}_2\left(\frac{1}{\Lambda  \tau  R+R}\right) (\Lambda  \tau +(\Lambda  \tau -1) \log (R)-2)}{(\Lambda  \tau
   +1)^3} +\frac{2 (\Lambda  \tau -1) \log ^3(R (\Lambda  \tau +1))}{3 (\Lambda  \tau
   +1)^3} \nonumber \\
   && \frac{2 \left(\pi ^2 (\Lambda  \tau -1)+\frac{3 \Lambda  \tau  ((\Lambda  \tau +1) \log (\Lambda  \tau +1)-\log
   (R)+2)}{1-(1+\Lambda   \tau) R}\right) \log (R (\Lambda  \tau +1))}{3 (\Lambda  \tau +1)^3}  \nonumber \\
    && \hspace{-2cm} -\frac{\log \left(\frac{1}{R (\Lambda  \tau +1)}\right) \left(2 \Lambda  \tau  (\Lambda  \tau -2)+(\Lambda  \tau
   -1)^2 (-\log (\Lambda  \tau +1))+\left(\Lambda ^2 \tau ^2-1\right) \log (R)\right) \log \left(1-\frac{1}{\Lambda  R \tau
   +R}\right)}{(\Lambda  \tau +1)^3} \nonumber \, .
\eea
As stated in \Subsec{subsec:NHR}
\be
S_3=0\,.
\ee
However, to be able to compute $A_3$ and $B_3$ it is necessary to determine $S_4$, which is given by 
\be
S_4=\frac{2 \Lambda  \tau  \left(2 \text{Li}_3\left(\frac{1}{R (\Lambda  \tau +1)}\right)-\text{Li}_2\left(\frac{1}{R (\Lambda  \tau +1)}\right) \log \left(\frac{1}{R (\Lambda  \tau
   +1)}\right)\right)}{(\Lambda  \tau +1)^3} \,.
\ee

\newpage

\bibliography{LDBJbib}{}

\providecommand{\href}[2]{#2}\begingroup\raggedright\begin{thebibliography}{10}

\bibitem{CasalderreySolana:2011us}
J.~Casalderrey-Solana, H.~Liu, D.~Mateos, K.~Rajagopal and U.~A. Wiedemann,
  {\it {Gauge/String Duality, Hot QCD and Heavy Ion Collisions}},
  \href{http://arXiv.org/abs/1101.0618}{{\tt 1101.0618}}.
%%CITATION = ARXIV:1101.0618;%%

\bibitem{Hartnoll:2016apf}
S.~A. Hartnoll, A.~Lucas and S.~Sachdev, {\it {Holographic quantum matter}},
  \href{http://arXiv.org/abs/1612.07324}{{\tt 1612.07324}}.
%%CITATION = ARXIV:1612.07324;%%

\bibitem{Chesler:2013lia}
P.~M. Chesler and L.~G. Yaffe, {\it {Numerical solution of gravitational
  dynamics in asymptotically anti-de Sitter spacetimes}},  {\em JHEP} {\bf 07}
  (2014) 086 [\href{http://arXiv.org/abs/1309.1439}{{\tt 1309.1439}}].
%%CITATION = ARXIV:1309.1439;%%

\bibitem{Emparan:2013moa}
R.~Emparan, R.~Suzuki and K.~Tanabe, {\it {The large D limit of General
  Relativity}},  {\em JHEP} {\bf 06} (2013) 009
  [\href{http://arXiv.org/abs/1302.6382}{{\tt 1302.6382}}].
%%CITATION = ARXIV:1302.6382;%%

\bibitem{Emparan:2013xia}
R.~Emparan, D.~Grumiller and K.~Tanabe, {\it {Large-D gravity and low-D
  strings}},  {\em Phys. Rev. Lett.} {\bf 110} (2013), no.~25 251102
  [\href{http://arXiv.org/abs/1303.1995}{{\tt 1303.1995}}].
%%CITATION = ARXIV:1303.1995;%%

\bibitem{Emparan:2013oza}
R.~Emparan and K.~Tanabe, {\it {Holographic superconductivity in the large D
  expansion}},  {\em JHEP} {\bf 01} (2014) 145
  [\href{http://arXiv.org/abs/1312.1108}{{\tt 1312.1108}}].
%%CITATION = ARXIV:1312.1108;%%

\bibitem{Romero-Bermudez:2015bma}
A.~M. Garc\'{i}a-Garc\'{i}a and A.~Romero-Berm\'{u}dez, {\it {Conductivity and
  entanglement entropy of high dimensional holographic superconductors}},  {\em
  JHEP} {\bf 09} (2015) 033 [\href{http://arXiv.org/abs/1502.03616}{{\tt
  1502.03616}}].
%%CITATION = ARXIV:1502.03616;%%

\bibitem{Emparan:2015rva}
R.~Emparan, R.~Suzuki and K.~Tanabe, {\it {Quasinormal modes of (Anti-)de
  Sitter black holes in the 1/D expansion}},  {\em JHEP} {\bf 04} (2015) 085
  [\href{http://arXiv.org/abs/1502.02820}{{\tt 1502.02820}}].
%%CITATION = ARXIV:1502.02820;%%

\bibitem{Andrade:2015hpa}
T.~Andrade, S.~A. Gentle and B.~Withers, {\it {Drude in D major}},  {\em JHEP}
  {\bf 06} (2016) 134 [\href{http://arXiv.org/abs/1512.06263}{{\tt
  1512.06263}}].
%%CITATION = ARXIV:1512.06263;%%

\bibitem{Herzog:2016hob}
C.~P. Herzog, M.~Spillane and A.~Yarom, {\it {The holographic dual of a Riemann
  problem in a large number of dimensions}},  {\em JHEP} {\bf 08} (2016) 120
  [\href{http://arXiv.org/abs/1605.01404}{{\tt 1605.01404}}].
%%CITATION = ARXIV:1605.01404;%%

\bibitem{Rozali:2017bll}
M.~Rozali, E.~Sabag and A.~Yarom, {\it {Holographic Turbulence in a Large
  Number of Dimensions}},  {\em JHEP} {\bf 04} (2018) 065
  [\href{http://arXiv.org/abs/1707.08973}{{\tt 1707.08973}}].
%%CITATION = ARXIV:1707.08973;%%

\bibitem{Emparan:2016sjk}
R.~Emparan, K.~Izumi, R.~Luna, R.~Suzuki and K.~Tanabe, {\it {Hydro-elastic
  Complementarity in Black Branes at large D}},  {\em JHEP} {\bf 06} (2016) 117
  [\href{http://arXiv.org/abs/1602.05752}{{\tt 1602.05752}}].
%%CITATION = ARXIV:1602.05752;%%

\bibitem{Iizuka:2018zgt}
N.~Iizuka, A.~Ishibashi and K.~Maeda, {\it {Cosmic Censorship at Large D:
  Stability analysis in polarized AdS black branes (holes)}},  {\em JHEP} {\bf
  03} (2018) 177 [\href{http://arXiv.org/abs/1801.07268}{{\tt 1801.07268}}].
%%CITATION = ARXIV:1801.07268;%%

\bibitem{Tanabe:2015hda}
K.~Tanabe, {\it {Black rings at large D}},  {\em JHEP} {\bf 02} (2016) 151
  [\href{http://arXiv.org/abs/1510.02200}{{\tt 1510.02200}}].
%%CITATION = ARXIV:1510.02200;%%

\bibitem{Bhattacharyya:2015fdk}
S.~Bhattacharyya, M.~Mandlik, S.~Minwalla and S.~Thakur, {\it {A Charged
  Membrane Paradigm at Large D}},  {\em JHEP} {\bf 04} (2016) 128
  [\href{http://arXiv.org/abs/1511.03432}{{\tt 1511.03432}}].
%%CITATION = ARXIV:1511.03432;%%

\bibitem{Tanabe:2016opw}
K.~Tanabe, {\it {Charged rotating black holes at large D}},
  \href{http://arXiv.org/abs/1605.08854}{{\tt 1605.08854}}.
%%CITATION = ARXIV:1605.08854;%%

\bibitem{Rozali:2016yhw}
M.~Rozali and A.~Vincart-Emard, {\it {On Brane Instabilities in the Large $D$
  Limit}},  {\em JHEP} {\bf 08} (2016) 166
  [\href{http://arXiv.org/abs/1607.01747}{{\tt 1607.01747}}].
%%CITATION = ARXIV:1607.01747;%%

\bibitem{Dandekar:2016fvw}
Y.~Dandekar, A.~De, S.~Mazumdar, S.~Minwalla and A.~Saha, {\it {The large D
  black hole Membrane Paradigm at first subleading order}},  {\em JHEP} {\bf
  12} (2016) 113 [\href{http://arXiv.org/abs/1607.06475}{{\tt 1607.06475}}].
%%CITATION = ARXIV:1607.06475;%%

\bibitem{Dandekar:2016jrp}
Y.~Dandekar, S.~Mazumdar, S.~Minwalla and A.~Saha, {\it {Unstable `black
  branes' from scaled membranes at large $D$}},  {\em JHEP} {\bf 12} (2016) 140
  [\href{http://arXiv.org/abs/1609.02912}{{\tt 1609.02912}}].
%%CITATION = ARXIV:1609.02912;%%

\bibitem{Bhattacharyya:2016nhn}
S.~Bhattacharyya, A.~K. Mandal, M.~Mandlik, U.~Mehta, S.~Minwalla, U.~Sharma
  and S.~Thakur, {\it {Currents and Radiation from the large $D$ Black Hole
  Membrane}},  {\em JHEP} {\bf 05} (2017) 098
  [\href{http://arXiv.org/abs/1611.09310}{{\tt 1611.09310}}].
%%CITATION = ARXIV:1611.09310;%%

\bibitem{Chen:2017wpf}
B.~Chen, P.-C. Li and Z.-z. Wang, {\it {Charged Black Rings at large D}},  {\em
  JHEP} {\bf 04} (2017) 167 [\href{http://arXiv.org/abs/1702.00886}{{\tt
  1702.00886}}].
%%CITATION = ARXIV:1702.00886;%%

\bibitem{Chen:2017hwm}
B.~Chen and P.-C. Li, {\it {Static Gauss-Bonnet Black Holes at Large $D$}},
  {\em JHEP} {\bf 05} (2017) 025 [\href{http://arXiv.org/abs/1703.06381}{{\tt
  1703.06381}}].
%%CITATION = ARXIV:1703.06381;%%

\bibitem{Bhattacharyya:2017hpj}
S.~Bhattacharyya, P.~Biswas, B.~Chakrabarty, Y.~Dandekar and A.~Dinda, {\it
  {The large D black hole dynamics in AdS/dS backgrounds}},
  \href{http://arXiv.org/abs/1704.06076}{{\tt 1704.06076}}.
%%CITATION = ARXIV:1704.06076;%%

\bibitem{Miyamoto:2017ozn}
U.~Miyamoto, {\it {Non-linear perturbation of black branes at large $D$}},
  {\em JHEP} {\bf 06} (2017) 033 [\href{http://arXiv.org/abs/1705.00486}{{\tt
  1705.00486}}].
%%CITATION = ARXIV:1705.00486;%%

\bibitem{Chen:2017rxa}
B.~Chen, P.-C. Li and C.-Y. Zhang, {\it {Einstein-Gauss-Bonnet Black Strings at
  Large $D$}},  {\em JHEP} {\bf 10} (2017) 123
  [\href{http://arXiv.org/abs/1707.09766}{{\tt 1707.09766}}].
%%CITATION = ARXIV:1707.09766;%%

\bibitem{Herzog:2017qwp}
C.~P. Herzog and Y.~Kim, {\it {The Large Dimension Limit of a Small Black Hole
  Instability in Anti-de Sitter Space}},  {\em JHEP} {\bf 02} (2018) 167
  [\href{http://arXiv.org/abs/1711.04865}{{\tt 1711.04865}}].
%%CITATION = ARXIV:1711.04865;%%

\bibitem{Dandekar:2017aiv}
Y.~Dandekar, S.~Kundu, S.~Mazumdar, S.~Minwalla, A.~Mishra and A.~Saha, {\it
  {An Action for and Hydrodynamics from the improved Large D membrane}},
  \href{http://arXiv.org/abs/1712.09400}{{\tt 1712.09400}}.
%%CITATION = ARXIV:1712.09400;%%

\bibitem{Emparan:2018bmi}
R.~Emparan, R.~Luna, M.~Mart\'{i}nez, R.~Suzuki and K.~Tanabe, {\it {Phases and
  Stability of Non-Uniform Black Strings}},  {\em JHEP} {\bf 05} (2018) 104
  [\href{http://arXiv.org/abs/1802.08191}{{\tt 1802.08191}}].
%%CITATION = ARXIV:1802.08191;%%

\bibitem{Andrade:2018zeb}
T.~Andrade, C.~Pantelidou and B.~Withers, {\it {Large D holography with metric
  deformations}},  \href{http://arXiv.org/abs/1806.00306}{{\tt 1806.00306}}.
%%CITATION = ARXIV:1806.00306;%%

\bibitem{Andrade:2018nsz}
T.~Andrade, R.~Emparan and D.~Licht, {\it {Rotating black holes and black bars
  at large D}},  {\em JHEP} {\bf 09} (2018) 107
  [\href{http://arXiv.org/abs/1807.01131}{{\tt 1807.01131}}].
%%CITATION = ARXIV:1807.01131;%%

\bibitem{Bjorken:1982qr}
J.~D. Bjorken, {\it {Highly Relativistic Nucleus-Nucleus Collisions: The
  Central Rapidity Region}},  {\em Phys. Rev.} {\bf D27} (1983) 140--151.
%%CITATION = PHRVA,D27,140;%%

\bibitem{Janik:2005zt}
R.~A. Janik and R.~B. Peschanski, {\it {Asymptotic perfect fluid dynamics as a
  consequence of Ads/CFT}},  {\em Phys. Rev.} {\bf D73} (2006) 045013
  [\href{http://arXiv.org/abs/hep-th/0512162}{{\tt hep-th/0512162}}].
%%CITATION = HEP-TH/0512162;%%

\bibitem{Heller:2013fn}
M.~P. Heller, R.~A. Janik and P.~Witaszczyk, {\it {Hydrodynamic Gradient
  Expansion in Gauge Theory Plasmas}},  {\em Phys. Rev. Lett.} {\bf 110}
  (2013), no.~21 211602 [\href{http://arXiv.org/abs/1302.0697}{{\tt
  1302.0697}}].
%%CITATION = ARXIV:1302.0697;%%

\bibitem{Heller:2015dha}
M.~P. Heller and M.~Spalinski, {\it {Hydrodynamics Beyond the Gradient
  Expansion: Resurgence and Resummation}},  {\em Phys. Rev. Lett.} {\bf 115}
  (2015), no.~7 072501 [\href{http://arXiv.org/abs/1503.07514}{{\tt
  1503.07514}}].
%%CITATION = ARXIV:1503.07514;%%

\bibitem{Florkowski:2017olj}
W.~Florkowski, M.~P. Heller and M.~Spalinski, {\it {New theories of
  relativistic hydrodynamics in the LHC era}},  {\em Rept. Prog. Phys.} {\bf
  81} (2018), no.~4 046001 [\href{http://arXiv.org/abs/1707.02282}{{\tt
  1707.02282}}].
%%CITATION = ARXIV:1707.02282;%%

\bibitem{Romatschke:2017vte}
P.~Romatschke, {\it {Far From Equilibrium Fluid Dynamics}},
  \href{http://arXiv.org/abs/1704.08699}{{\tt 1704.08699}}.
%%CITATION = ARXIV:1704.08699;%%

\bibitem{Denicol:2017lxn}
G.~S. Denicol and J.~Noronha, {\it {Analytical attractor and the divergence of
  the slow-roll expansion in relativistic hydrodynamics}},
  \href{http://arXiv.org/abs/1711.01657}{{\tt 1711.01657}}.
%%CITATION = ARXIV:1711.01657;%%

\bibitem{Strickland:2017kux}
M.~Strickland, J.~Noronha and G.~Denicol, {\it {The anisotropic non-equilibrium
  hydrodynamic attractor}},  \href{http://arXiv.org/abs/1709.06644}{{\tt
  1709.06644}}.
%%CITATION = ARXIV:1709.06644;%%

\bibitem{Spalinski:2017mel}
M.~Spali\'nski, {\it {On the hydrodynamic attractor of Yang-Mills plasma}},
  \href{http://arXiv.org/abs/1708.01921}{{\tt 1708.01921}}.
%%CITATION = ARXIV:1708.01921;%%

\bibitem{Romatschke:2017acs}
P.~Romatschke, {\it {Relativistic Hydrodynamic Attractors with Broken
  Symmetries: Non-Conformal and Non-Homogeneous}},
  \href{http://arXiv.org/abs/1710.03234}{{\tt 1710.03234}}.
%%CITATION = ARXIV:1710.03234;%%

\bibitem{Heller:2016rtz}
M.~P. Heller, A.~Kurkela and M.~Spalinski, {\it {Hydrodynamization and
  transient modes of expanding plasma in kinetic theory}},
  \href{http://arXiv.org/abs/1609.04803}{{\tt 1609.04803}}.
%%CITATION = ARXIV:1609.04803;%%

\bibitem{Casalderrey-Solana:2017zyh}
J.~Casalderrey-Solana, N.~I. Gushterov and B.~Meiring, {\it {Resurgence and
  Hydrodynamic Attractors in Gauss-Bonnet Holography}},  {\em JHEP} {\bf 04}
  (2018) 042 [\href{http://arXiv.org/abs/1712.02772}{{\tt 1712.02772}}].
%%CITATION = ARXIV:1712.02772;%%

\bibitem{Behtash:2018moe}
A.~Behtash, S.~Kamata, M.~Martinez and C.~N. Cruz-Camacho, {\it
  {Non-perturbative rheological behavior of a far-from-equilibrium expanding
  plasma}},  \href{http://arXiv.org/abs/1805.07881}{{\tt 1805.07881}}.
%%CITATION = ARXIV:1805.07881;%%

\bibitem{Denicol:2018pak}
G.~S. Denicol and J.~Noronha, {\it {Hydrodynamic attractor and the fate of
  perturbative expansions in Gubser flow}},
  \href{http://arXiv.org/abs/1804.04771}{{\tt 1804.04771}}.
%%CITATION = ARXIV:1804.04771;%%

\bibitem{Heller:2018qvh}
M.~P. Heller and V.~Svensson, {\it {How does relativistic kinetic theory
  remember about initial conditions?}},
  \href{http://arXiv.org/abs/1802.08225}{{\tt 1802.08225}}.
%%CITATION = ARXIV:1802.08225;%%

\bibitem{Behtash:2017wqg}
A.~Behtash, C.~N. Cruz-Camacho and M.~Martinez, {\it {Far-from-equilibrium
  attractors and nonlinear dynamical systems approach to the Gubser flow}},
  {\em Phys. Rev.} {\bf D97} (2018), no.~4 044041
  [\href{http://arXiv.org/abs/1711.01745}{{\tt 1711.01745}}].
%%CITATION = ARXIV:1711.01745;%%

\bibitem{Baier:2007ix}
R.~Baier, P.~Romatschke, D.~T. Son, A.~O. Starinets and M.~A. Stephanov, {\it
  {Relativistic viscous hydrodynamics, conformal invariance, and holography}},
  {\em JHEP} {\bf 04} (2008) 100 [\href{http://arXiv.org/abs/0712.2451}{{\tt
  0712.2451}}].
%%CITATION = ARXIV:0712.2451;%%

\bibitem{Kovtun:2004de}
P.~Kovtun, D.~T. Son and A.~O. Starinets, {\it {Viscosity in strongly
  interacting quantum field theories from black hole physics}},  {\em Phys.
  Rev. Lett.} {\bf 94} (2005) 111601
  [\href{http://arXiv.org/abs/hep-th/0405231}{{\tt hep-th/0405231}}].
%%CITATION = HEP-TH/0405231;%%

\bibitem{Bhattacharyya:2008mz}
S.~Bhattacharyya, R.~Loganayagam, I.~Mandal, S.~Minwalla and A.~Sharma, {\it
  {Conformal Nonlinear Fluid Dynamics from Gravity in Arbitrary Dimensions}},
  {\em JHEP} {\bf 12} (2008) 116 [\href{http://arXiv.org/abs/0809.4272}{{\tt
  0809.4272}}].
%%CITATION = ARXIV:0809.4272;%%

\bibitem{Heller:2011ju}
M.~P. Heller, R.~A. Janik and P.~Witaszczyk, {\it {The characteristics of
  thermalization of boost-invariant plasma from holography}},  {\em Phys. Rev.
  Lett.} {\bf 108} (2012) 201602 [\href{http://arXiv.org/abs/1103.3452}{{\tt
  1103.3452}}].
%%CITATION = ARXIV:1103.3452;%%

\bibitem{Heller:2012je}
M.~P. Heller, R.~A. Janik and P.~Witaszczyk, {\it {A numerical relativity
  approach to the initial value problem in asymptotically Anti-de Sitter
  spacetime for plasma thermalization - an ADM formulation}},  {\em Phys. Rev.}
  {\bf D85} (2012) 126002 [\href{http://arXiv.org/abs/1203.0755}{{\tt
  1203.0755}}].
%%CITATION = ARXIV:1203.0755;%%

\bibitem{Chesler:2009cy}
P.~M. Chesler and L.~G. Yaffe, {\it {Boost invariant flow, black hole
  formation, and far-from-equilibrium dynamics in N = 4 supersymmetric
  Yang-Mills theory}},  {\em Phys. Rev.} {\bf D82} (2010) 026006
  [\href{http://arXiv.org/abs/0906.4426}{{\tt 0906.4426}}].
%%CITATION = ARXIV:0906.4426;%%

\bibitem{Bhattacharyya:2008jc}
S.~Bhattacharyya, V.~E. Hubeny, S.~Minwalla and M.~Rangamani, {\it {Nonlinear
  Fluid Dynamics from Gravity}},  {\em JHEP} {\bf 02} (2008) 045
  [\href{http://arXiv.org/abs/0712.2456}{{\tt 0712.2456}}].
%%CITATION = ARXIV:0712.2456;%%

\bibitem{Kinoshita:2008dq}
S.~Kinoshita, S.~Mukohyama, S.~Nakamura and K.-y. Oda, {\it {A Holographic Dual
  of Bjorken Flow}},  {\em Prog. Theor. Phys.} {\bf 121} (2009) 121--164
  [\href{http://arXiv.org/abs/0807.3797}{{\tt 0807.3797}}].
%%CITATION = ARXIV:0807.3797;%%

\bibitem{Bhattacharyya:2018iwt}
S.~Bhattacharyya, P.~Biswas and M.~Patra, {\it {A leading-order comparison
  between fluid-gravity and membrane-gravity dualities}},
  \href{http://arXiv.org/abs/1807.05058}{{\tt 1807.05058}}.
%%CITATION = ARXIV:1807.05058;%%

\bibitem{Emparan:2014cia}
R.~Emparan and K.~Tanabe, {\it {Universal quasinormal modes of large D black
  holes}},  {\em Phys. Rev.} {\bf D89} (2014), no.~6 064028
  [\href{http://arXiv.org/abs/1401.1957}{{\tt 1401.1957}}].
%%CITATION = ARXIV:1401.1957;%%

\bibitem{Emparan:2014aba}
R.~Emparan, R.~Suzuki and K.~Tanabe, {\it {Decoupling and non-decoupling
  dynamics of large D black holes}},  {\em JHEP} {\bf 07} (2014) 113
  [\href{http://arXiv.org/abs/1406.1258}{{\tt 1406.1258}}].
%%CITATION = ARXIV:1406.1258;%%

\bibitem{Basar:2015ava}
G.~Basar and G.~V. Dunne, {\it {Hydrodynamics, resurgence, and
  transasymptotics}},  {\em Phys. Rev.} {\bf D92} (2015), no.~12 125011
  [\href{http://arXiv.org/abs/1509.05046}{{\tt 1509.05046}}].
%%CITATION = ARXIV:1509.05046;%%

\bibitem{Spalinski:2018mqg}
M.~Spalinski, {\it {Universal behaviour, transients and attractors in
  supersymmetric Yang-Mills plasma}},
  \href{http://arXiv.org/abs/1805.11689}{{\tt 1805.11689}}.
%%CITATION = ARXIV:1805.11689;%%

\bibitem{Aniceto:2018uik}
I.~Aniceto, B.~Meiring, J.~Jankowski and M.~Spalinski, {\it {The large
  proper-time expansion of Yang-Mills plasma as a resurgent transseries}},
  \href{http://arXiv.org/abs/1810.07130}{{\tt 1810.07130}}.
%%CITATION = ARXIV:1810.07130;%%

\bibitem{Grozdanov:2015kqa}
S.~Grozdanov and N.~Kaplis, {\it {Constructing higher-order hydrodynamics: The
  third order}},  {\em Phys. Rev.} {\bf D93} (2016), no.~6 066012
  [\href{http://arXiv.org/abs/1507.02461}{{\tt 1507.02461}}].
%%CITATION = ARXIV:1507.02461;%%

\bibitem{Kovtun:2005ev}
P.~K. Kovtun and A.~O. Starinets, {\it {Quasinormal modes and holography}},
  {\em Phys. Rev.} {\bf D72} (2005) 086009
  [\href{http://arXiv.org/abs/hep-th/0506184}{{\tt hep-th/0506184}}].
%%CITATION = HEP-TH/0506184;%%

\bibitem{Natario:2004jd}
J.~Natario and R.~Schiappa, {\it {On the classification of asymptotic
  quasinormal frequencies for d-dimensional black holes and quantum gravity}},
  {\em Adv. Theor. Math. Phys.} {\bf 8} (2004), no.~6 1001--1131
  [\href{http://arXiv.org/abs/hep-th/0411267}{{\tt hep-th/0411267}}].
%%CITATION = HEP-TH/0411267;%%

\bibitem{Betzios:2018kwn}
P.~Betzios, U.~Gursoy, M.~Jarvinen and G.~Policastro, {\it {Fluctuations in
  non-conformal holographic plasma at criticality}},
  \href{http://arXiv.org/abs/1807.01718}{{\tt 1807.01718}}.
%%CITATION = ARXIV:1807.01718;%%

\end{thebibliography}\endgroup
\bibliographystyle{JHEP-2}
\end{document}